\documentclass[
aps,
 ,twocolumn
]{revtex4}
\usepackage{graphicx}
\usepackage[caption=false]{subfig}
\usepackage{amsmath}
\usepackage{amssymb}
\usepackage{color}
\usepackage{hyperref}
\newcommand{\bs}[1]{\boldsymbol{#1}}
\hypersetup{colorlinks,linkcolor={blue},citecolor={red},urlcolor={blue}}
\usepackage{qcircuit}
\usepackage{epstopdf}

\newcommand{\ket}[1]{| #1 \rangle}
\newcommand{\bra}[1]{ \langle  #1 |}
\newcommand{\braket}[2]{\langle #1 | #2 \rangle}

\begin{document}

\title{Implementing quantum algorithms on  temporal photonic cluster states}

\author{Daiqin Su}
\email{daiqin@xanadu.ai}
\author{Krishna Kumar Sabapathy}
\author{Casey R. Myers}
\author{Haoyu Qi}
\author{Christian Weedbrook}
\author{Kamil Br\'adler}
\affiliation{Xanadu, 372 Richmond Street West, Toronto, Ontario M5V 1X6, Canada }



%
\begin{abstract}
{ Implementing quantum algorithms is essential for quantum computation. 
We study the implementation of three quantum algorithms by performing homodyne measurements on a two-dimensional temporal continuous-variable cluster state. We first review the generation of temporal cluster states and the implementation of gates using the measurement-based model. Alongside this we discuss methods to introduce non-Gaussianity into the cluster states. The first algorithm we consider is Gaussian Boson Sampling in which only Gaussian unitaries need to be implemented. Taking into account the fact that 
input states are also Gaussian, the errors due to the effect of finite squeezing can be corrected, provided a moderate amount of online squeezing
is available. This helps to construct a large Gaussian Boson Sampling machine. The second algorithm is the continuous-variable Instantaneous
Quantum Polynomial circuit in which one needs to implement non-Gaussian gates, such as the cubic phase gate. We discuss several methods of
implementing the cubic phase gate and fit them into the temporal cluster state architecture.
The third algorithm is the continuous-variable version of Grover's search algorithm, the main challenge of which is the implementation of the inversion operator. We propose a method to implement the inversion operator
by injecting a resource state into a teleportation circuit. The resource state is simulated using the Strawberry Fields quantum software package.}
\end{abstract}

\maketitle

\tableofcontents

\section{Introduction}

Measurement-based (or one way) quantum computation is a particular model of quantum computation \cite{Raussendorf2001}. It is based on a multipartite entangled resource state called a 
cluster state \cite{Briegel2001}, and local measurements. For continuous-variable (CV) measurement-based quantum computation \cite{Menicucci2006, Loock2007example, Gu2009}, the cluster state
is a highly entangled multimode Gaussian state and the required local measurements are homodyne and non-Gaussian measurements. 
One of the main challenges of 
measurement-based quantum computation is to generate a scalable and universal cluster state. Several ways of generating CV cluster states
have been proposed \cite{Zhang2006, Loock2007, Menicucci2008, Menicucci2010, Menicucci2011} and some of them have been experimentally realised \cite{Yokoyama2013, Yoshikawa2016, Chen2014}. In particular, the temporal CV cluster state architecture is advantageous in terms of
the scalability \cite{Alexander2018Universal} because it requires only a small number of optical elements. 
A one-dimensional temporal CV cluster state with 10,000 entangled modes \cite{Yokoyama2013}, as well as a one-million-mode version \cite{Yoshikawa2016}, 
have been experimentally generated. However, measuring the one-dimensional cluster states can only implement single-mode unitaries. 
To implement arbitrary unitaries, two-dimensional 
temporal cluster states are required \cite{Menicucci2011}. A method to generate two-dimensional temporal cluster states has been proposed by Menicucci \cite{Menicucci2011}.
Given the successful generation of the large one-dimensional temporal cluster states, the experimental realisation of two-dimensional 
temporal cluster states can be expected in the near future. 

Our work in this paper is based on two-dimensional temporal cluster states. 
The implementation of a set of universal gates (phase shift, squeezing gate, cubic phase gate, beam splitter, {\rm etc.}) via homodyne measurements and
non-Gaussian resource states on a two-dimensional temporal cluster state were discussed in Refs. \cite{Alexander2017MBLO, Alexander2018Universal}. 
This constitutes the first step towards a universal measurement-based
quantum computation. The natural next step is to implement some particular algorithms based on this set of universal gates. In this work we focus on 
implementing three important quantum algorithms: Gaussian Boson Sampling \cite{Hamilton2017}, 
continuous-variable Instantaneous Quantum Polynomial (CV-IQP) circuit \cite{Douce2017CVIQP} and the CV Grover's search algorithm \cite{Pati2000}. 

In Gaussian Boson Sampling a set of squeezed states are injected into a linear multimode interferometer and the output state is measured by 
photon number resolved detectors (PNR) to obtain the photon statistics. It is evident that only Gaussian states and Gaussian unitaries are involved, 
and the only non-Gaussian element is the photon number detection. 
This makes the implementation relatively easy: only Gaussian gates are required. In addition, for the Gaussian unitaries and Gaussian states the errors due to the effect of finite squeezing can be corrected \cite{Su2018EC}, provided a moderate amount of online squeezing is available. 
It is therefore possible, in principle, to conceive a Gaussian Boson Sampling machine with a large number of modes. 

For the CV-IQP circuit \cite{Douce2017CVIQP}, non-Gaussian gates are required. In particular, we consider commuting unitaries that  are functions of the position quadrature. The lowest order non-Gaussian gate is the cubic phase gate, which has been studied extensively \cite{Krishna2018ON}. We summarise various implementations of  the cubic phase gate and explore which are better suited for measurement-based quantum computation.  The direct implementation of higher-order non-Gaussian gates is more challenging. However, they can be decomposed into cubic phase gates and Gaussian gates \cite{BraunsteinLloyd1999}.

The continuous-variable Grover's search algorithm \cite{Grover1997, Pati2000} is another algorithm that requires non-Gaussian gates. In this case the so-called ``Grover diffusion operator'' is challenging and a direct discrete-variable analog cannot be used. Instead this operator must be implemented via non-Gaussian gate teleportation. We consider two methods to implement the algorithm logic, using one and two continuous variable qumodes. We show in both cases that the Grover diffusion operator reduces to a sequence of higher-order quadrature phase gates. 


The paper is organised as follows: 
in Sec \ref{sec:1Dcluster}, we review the 
generation of one-dimensional temporal cluster states and the implementation of single-mode unitaries. In particular, we focus on the implementation of the cubic phase gate. We also discuss several methods to introduce non-Gaussianity into the temporal cluster state.
In Sec. \ref{sec:2Dcluster}, we summarise the generation of the universal two-dimensional cluster states and the implementation of two-mode unitaries, such as the beam splitter.
Sec. \ref{sec:GBS} discusses the implementation of the Gaussian Boson Sampling, Sec. \ref{sec:IQP} discusses the implementation of CV-IQP and 
Sec. \ref{sec:CVGrover} discusses the implementation of CV Grover's search algorithm. We conclude in Sec. \ref{sec:conclusion}.

\section{One-dimensional temporal cluster state}\label{sec:1Dcluster}

\subsection{Gate teleportation and basic elements of graphical representation}\label{sec:gate-teleportation}

\begin{figure}[ht!]
\includegraphics[width=8.5cm]{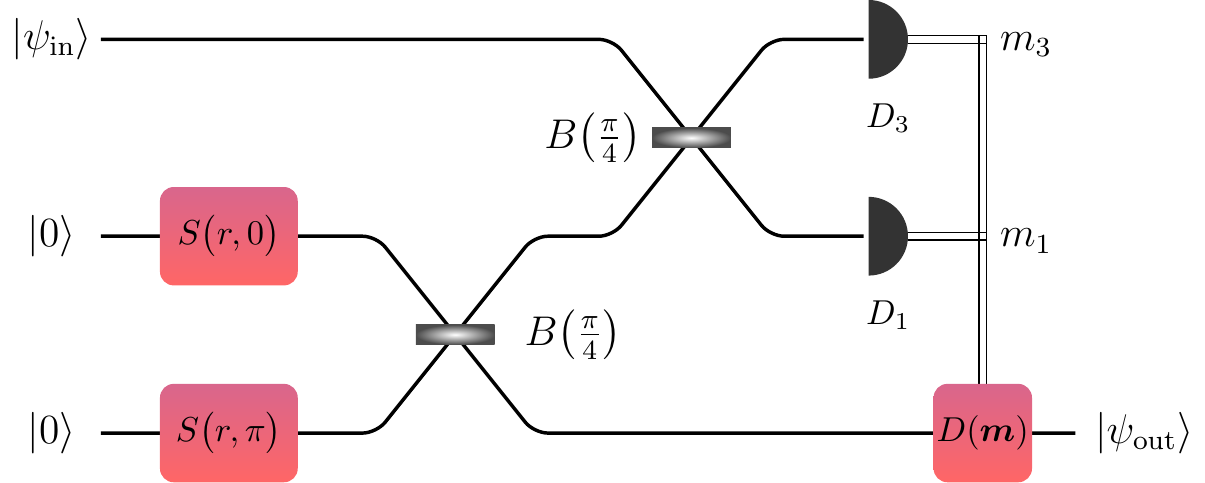}
\caption{Gate teleportation circuit. Two single-mode squeezed vacuum states are generated by squeezing the vacuum using two single-mode squeezers, 
the action of which is represented by the single-mode squeezing operator $S (r, \theta)$ with $\theta= 0$ and $\pi$, respectively. A two-mode squeezed vacuum 
state is produced after the two single-mode squeezed states pass through a beam splitter $B(\frac{\pi}{4})$ (a $50:50$ beam splitter). The input mode (with input state $| \psi_{\rm in} \rangle$) couples with
one of the two modes of the two-mode squeezed state via a beam splitter $B(\frac{\pi}{4})$, the outputs of which are detected by two homodyne detectors $D_1$ and $D_3$. 
The homodyne measurement outcomes ${m}_1$ and ${m}_3$ are used to displace the output state in the other mode of the two-mode squeezed state. The displacement operator
is denoted as $D (\boldsymbol m)$ where $\boldsymbol m= (m_1, m_3)$. A unitary is implemented on the input state that depends on the measurement quadratures of the homodyne detectors. } 
\label{fig:gate-teleportation}
\end{figure}

CV quantum teleportation \cite{Pirandola2015}, or CV gate teleportation \cite{Weedbrook2012RMP}, 
is the fundamental building block of CV measurement-based quantum computation. To implement a 
measurement-based Gaussian unitary, the input mode is coupled with one of the two modes of a two-mode squeezed state, the outputs of which are detected with two 
homodyne detectors, as shown in Fig. \ref{fig:gate-teleportation}. The input state is then teleported to another mode of the two-mode squeezed state, 
with an additional Gaussian unitary acting on it. The implemented Gaussian unitary depends on the measurement quadratures of the
homodyne detection. By changing the measurement quadratures, an arbitrary single-mode Gaussian unitary can be implemented. Here the two-mode squeezed state 
plays the role of a resource state for gate teleportation. The implementation of non-Gaussian gates will be discussed in Sec. \ref{sec:cubic}. 

\begin{figure*}[ht!]
\includegraphics[width=14cm]{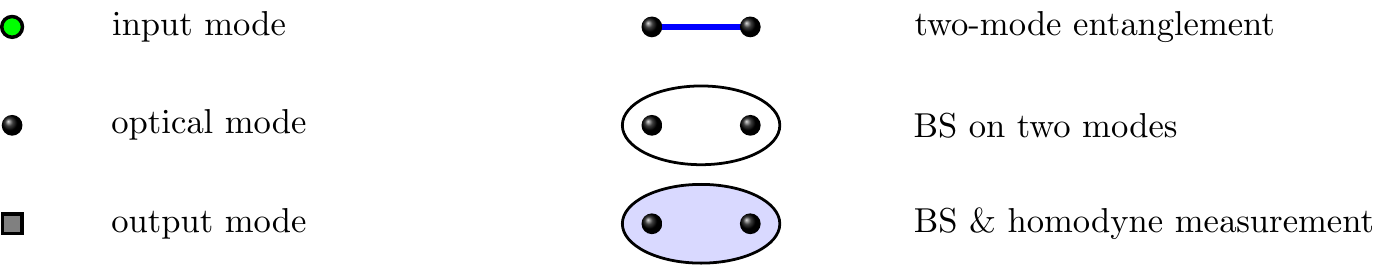}
\caption{ Elements of the simplified graphical representation of the CV cluster state and measurement-based quantum computation \cite{Alexander2016}. 
The shaded green dot represents an input mode, the solid black dot represents a general optical mode and
the grey square represents an output mode. A blue link between two optical modes represents two-mode entanglement. Two-mode entanglement is generated by applying 
a $50:50$ beam splitter (BS) to two squeezed pulses with orthogonal squeezing directions. An ellipse encircling two optical modes represents applying a $50:50$ 
beam splitter to them. An ellipse filled with light blue represents applying a $50:50$ beam splitter and then performing homodyne measurements. } 
\label{fig:graph-elements}
\end{figure*}

\begin{figure}[ht!]
\includegraphics[width=8.5cm]{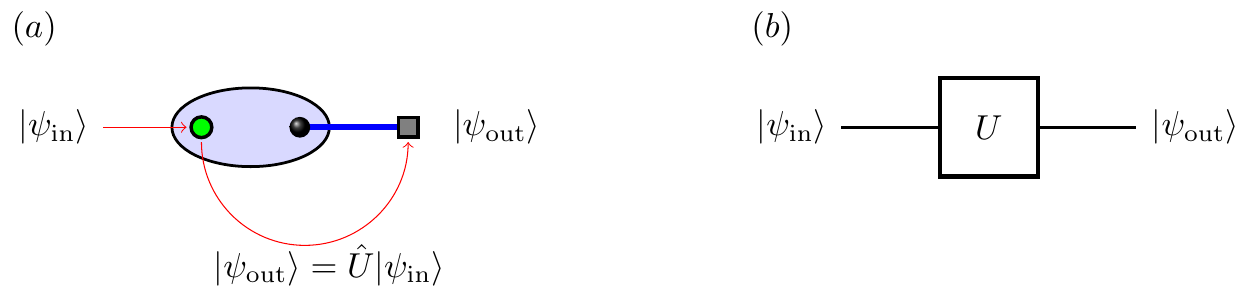}
\caption{ (a) Graphical representation of the gate teleportation shown in Fig. \ref{fig:gate-teleportation}. All relevant elements are introduced in Fig. \ref{fig:graph-elements}.
(b) An equivalent gate model circuit. $| \psi_{\rm in} \rangle$, $| \psi_{\rm out} \rangle$ and $\hat U$ are the input state, output state and implemented unitary respectively. } 
\label{fig:gate-teleportation-graph}
\end{figure}

Our main interest in this paper is the CV cluster state, which is the resource state for universal measurement-based quantum computation. By mentioning a CV cluster state we mean a
pure entangled multimode Gaussian state, although non-Gaussian CV cluster states are also available \cite{Walschaers2018nonGaussianCluster}, but are not conventional. 
It is convenient to represent CV cluster states graphically \cite{Menicucci2011graph}. It is also possible to represent the
measurements, input and output modes graphically. We follow Ref. \cite{Alexander2016} and introduce the graphical representation of the CV cluster state and 
measurement-based quantum computation. Fig. \ref{fig:graph-elements} shows some of the basic graph elements that we are going to use (more elements will be introduced
in the following sections). As an example, Fig. \ref{fig:gate-teleportation-graph} (a) shows a graphical representation of the gate teleportation in Fig. \ref{fig:gate-teleportation}, and
Fig. \ref{fig:gate-teleportation-graph} (b) shows the corresponding gate model circuit.

\subsection{Generation of one-dimensional temporal cluster state}

\begin{figure}[ht!]
\includegraphics[width=7.cm]{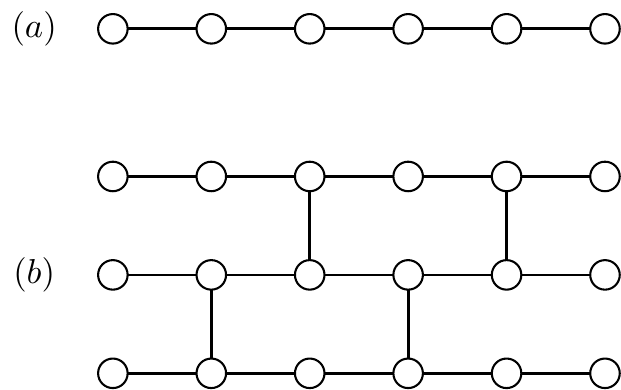}
\caption{ Standard CV cluster states: each circle (node) represents a single optical mode and the link between the modes represents the correlation (entanglement). 
We introduce different notations here to distinguish them from the temporal CV cluster states discussed in this paper. (a) One-dimensional cluster state. 
(b) Two-dimensional cluster state.  } 
\label{fig:standard-cluster-state}
\end{figure}

Fig. \ref{fig:standard-cluster-state}  shows the graphical representation of a standard one-dimensional 
and a two-dimensional CV cluster states: each node corresponds to a single optical mode and the links represent correlations between different optical modes. We use 
different notations in Fig. \ref{fig:standard-cluster-state}, as compared to Fig. \ref{fig:graph-elements}, to show the differences between the standard cluster states and 
the temporal cluster states discussed in this paper. 

\begin{figure*}[ht!]
\includegraphics[width=12cm]{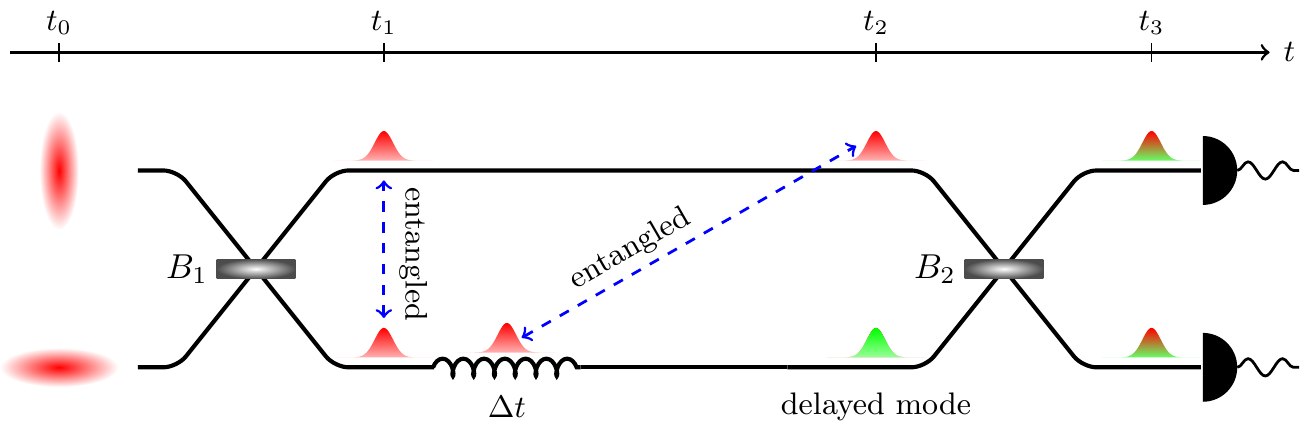}
\caption{ Optical setup that generates one-dimensional temporal cluster states \cite{Menicucci2011}. 
A series of pairs of single-mode squeezed pulses, with orthogonal squeezing directions,
are produced and injected into the optical setup with a repetition time interval $\Delta t$. The evolution of a single pair of single-mode squeezed pulses is illustrated. At time $t_0$, 
a pair of squeezed pulses are injected
into the setup. After passing through the first $50:50$ beam splitter $B_1$, a two-mode squeezed state is produced at $t_1$. The top mode keeps moving while the bottom mode
is delayed by the delay loop. When the top mode arrives at the second $50:50$ beam splitter $B_2$ at $t_2$, it couples with the delayed mode of an earlier pair of entangled 
modes. They hit the homodyne detectors at $t_3$. This process continues until we stop injecting squeezing pulses. In producing the temporal cluster states, 
we can choose not to detect the optical modes. } 
\label{fig:1D-cluster-setup}
\end{figure*}

\begin{figure}[ht!]
\includegraphics[width=8.5cm]{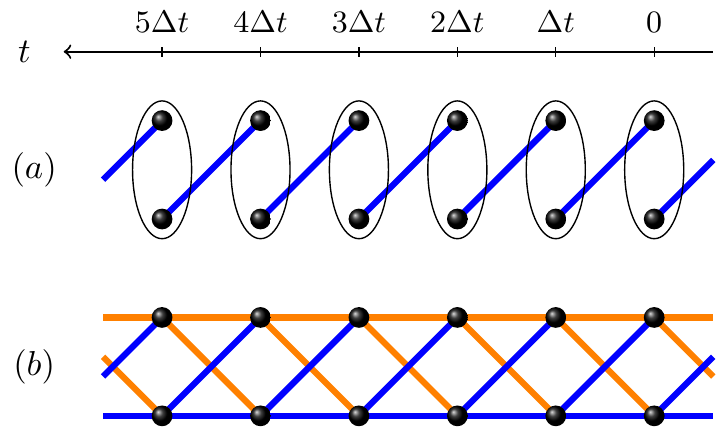}
\caption{ One-dimensional temporal cluster state that is produced by the optical setup in Fig. \ref{fig:1D-cluster-setup} \cite{Menicucci2011, Menicucci2011graph}.  (a) A series of two-mode squeezed vacuum states are 
generated with time interval $\Delta t$. One of the two modes of each two-mode squeezed state is delayed by $\Delta t$, and couples with the non-delayed mode of the latter
two-mode squeezed state via a $50:50$ beam splitter (represented by an ellipse). (b) Graphical representation of the one-dimensional temporal cluster state. Note that the
one-dimensional cluster state has a width of two nodes. We basically follow \cite{Menicucci2011} except colouring the nodes. The colour (blue and orange) 
of the links indicate the signs of the weights. } 
\label{fig:1D-cluster-state}
\end{figure}

In this section, we focus on the one-dimensional temporal CV cluster state \cite{Yokoyama2013}, 
which is generated by the optical setup shown in Fig. \ref{fig:1D-cluster-setup}. Optical parametric 
oscillators continuously generate pairs of single-mode squeezed vacuum states that are squeezed in orthogonal directions, e.g., one in position quadrature and the other in 
momentum quadrature. These squeezed pulses are then injected into the optical setup in Fig. \ref{fig:1D-cluster-setup}, 
which consists of two $50:50$ beam splitters and a delay loop \cite{Yokoyama2013}. 
The first beam splitter $B_1$ is used to generate a two-mode squeezed vacuum state. The delay loop delays the bottom mode by $\Delta t$, exactly the same as the time interval 
between two adjacent pairs of single-mode squeezed states. This is to make sure the top mode (non-delay mode) interferes with the delayed mode of an earlier two-mode 
squeezed state. If the output modes are not detected by the homodyne detectors, a temporal CV cluster state is generated, the graphical representation of which is shown in 
Fig. \ref{fig:1D-cluster-state}. In contrast to the spatial cluster states, the entanglement of the temporal cluster states is present between optical modes appearing at different times. 

\begin{figure*}[ht!]
\includegraphics[width=14cm]{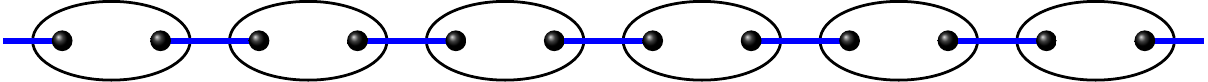}
\caption{ Simplified graphical representation of the one-dimensional temporal cluster state \cite{Alexander2016}. 
} 
\label{fig:1D-cluster-graph}
\end{figure*}

From Fig. \ref{fig:1D-cluster-state} (b) we see that the produced temporal cluster state has a width of two nodes. However, it still has dimension of one because it can only be used
to implement single-mode unitaries \cite{Menicucci2011}, as will be clear in the next section. To compare with the standard one-dimensional cluster states as in Fig. \ref{fig:standard-cluster-state} (a), and 
for ease of representation as the dimension increases, we will use a simplified graphical representation instead of ones as shown in Fig. \ref{fig:1D-cluster-state} (b). 
The one-dimensional temporal cluster state is represented by Fig. \ref{fig:1D-cluster-graph}, where an ellipse and the two optical modes that it encircles are together defined 
as a macronode \cite{Alexander2016}. The macronode can be considered as an analogue to the node in the standard cluster states shown in Fig. \ref{fig:standard-cluster-state}. 
However, one has to keep in mind that the ellipse represents a beam splitter transformation on the two modes, as defined in Fig. \ref{fig:graph-elements}. 

\subsection{Implementation of single-mode Gaussian gates}
In this subsection we recall the measurement-based implementation of Gaussian unitaries in a one-dimensional temporal cluster architecture \cite{Alexander2017MBLO, Alexander2016}. One can implement arbitrary single-mode Gaussian gates by directly choosing the measurement 
quadratures of the homodyne detection. 
Homodyne measurements $\hat{p}(\theta_1) = m_1$ and $\hat{p}(\theta_3)=m_3$, as shown in Fig. \ref{fig:gate-teleportation},  result in the implementation of the Gaussian unitary operator $\hat{A}(\theta_1,\theta_3,m_1,m_3)$ given by \cite{Alexander2016}
\begin{align}
&\hat{A}(\theta_1,\theta_3,m_1,m_3) \nonumber\\
&= D\left[-i\frac{e^{i\theta_1}m_3+e^{i\theta_3}m_1}{\sin(\theta_1-\theta_3)} \right] R(\theta_+) S({\rm ln} \tan{\theta_-}) R(\theta_+), 
\end{align}
where $S(r) = \exp[-r(\hat{a}^2-\hat{a}^{\dagger 2})/2]$, $R(\theta) = \exp(i\theta \hat{a}^{\dag}\hat{a})$, $D(\alpha) = \exp(\alpha \hat{a}^{\dag} - \alpha^* \hat{a})$, $\theta_{\pm} = (\theta_3\pm \theta_1)/2$. The operator is implemented on the immediate next macronode of the one-dimensional cluster that follows the macronode on which the measurement is performed. Note that while $\theta_1,\theta_3$ are the parameters under our control  that we can choose depending on the kind of unitary we want to implement, there is an additional displacement factor that is unavoidable and has to be accounted for in the feedforward. This covers all the Gaussian elements (up to displacement correction operators).

\subsection{Implementing single-mode non-Gaussian operations}\label{sec:cubic}
With regard to non-Gaussian elements, such as the cubic phase gate, one needs to use an additional optical setup where the non-Gaussianity is injected into the cluster through a resource state. This is achieved through a gate-teleportation circuit where the second homodyne detector in Fig. \ref{fig:gate-teleportation} (corresponding to outcome ${m}_1$) is replaced by an optical setup given by
\begin{align} \label{psub0} 
 \mbox{
\Qcircuit @C=0.7em @R=2.5em { 
\lstick{{\rm to~ cluster}}& \qw&\multigate{1}{B\left(\frac{\pi}{4}\right)} & \qw &\measureD{\mbox{$\Pi_{p_{\theta}}$}} & \cw ~~~~~~m_1\\ 
\lstick{\ket{\phi_r}}& \qw&\ghost{B\left(\frac{\pi}{4}\right)} & \qw &\measureD{\mbox{$\Pi_x$}} & \cw ~~~~~~~m_e\,.
}}
\end{align}
In Eq. \eqref{psub0}, $\ket{\phi_r}$ is any suitable resource state, $\Pi_x$ and $\Pi_{p_{\theta}}$ are homodyne measurements of the quadratures $\hat{x}$
 and $\hat{p}_{\theta} = \hat{p} \cos{\theta} + \hat{q} \sin{\theta}$. For a resource state  
\begin{align}
\ket{\phi_r} = \int \mathrm d x \phi_r(x) \ket{x} = \phi(\hat{x}) \ket{0}_p, 
\end{align} 
with $\ket{x}$ the position eigenstate and $\ket{0}_p =  \int \mathrm d x \ket{x}$ is the zero-momentum eigenstate. The outcome operator that applies to the next node of the cluster is given by \cite{Alexander2018Universal}
\begin{align}
\hat{L}(\phi_r,\theta,\bs{m}) 
& = Z([\sqrt{2} m_3 - m_e]/2) \hat{M}(\theta,m_1) \hat{\mathcal{E}}(\phi_r,m_e)\nonumber\\
&\quad\times X(-m_3) S(\ln{2}),\nonumber\\
\hat{M}(\theta,m_1) & =  X(-2m_1 \,{\rm sec}\theta) R(-\pi/2) S(-\ln{2}) P({\rm tan}\theta),\nonumber\\
\hat{\mathcal{E}}(\phi_r,m_e) & = \sqrt{2} X(-m_e) S(\ln{\sqrt{2}}) \phi_r(\sqrt{2}m_e-\hat{x}),
\label{e4}
\end{align}
where $\boldsymbol m = (m_3, m_1, m_e)$, and $X(s) = e^{-i s \hat p}$ and $Z(s) = e^{i s \hat x}$ are the displacement operators in $\hat x$ and $\hat p$, respectively. 
We now consider the simplest case where the resource state is an ideal cubic phase state and the implemented gate is the 
ideal cubic phase gate. \\

\noindent \textbf{Ideal cubic phase gate.} 
To achieve universal quantum computation using CV quantum systems, non-Gaussian gates are required \cite{BraunsteinLloyd1999}. A non-Gaussian gate corresponds to a Hamiltonian consisting of degree greater than quadratic in position and momentum quadrature operators (or annihilation and creation operators). The simplest and most widely used non-Gaussian gate 
is the cubic phase gate \cite{gkp}, which is defined as
\begin{eqnarray}
V (\gamma) =\exp \big(i \gamma \hat x^3/3 \big), 
\end{eqnarray}
where $\gamma$ is the gate strength. 
Correspondingly, we can define the idealised (and unnormalisable) cubic phase states by applying $\hat V (\gamma)$ to the zero-momentum eigenstate, namely,
\begin{eqnarray}
| \phi_{\gamma} \rangle = V (\gamma) \ket{0_p} = \int {\rm d} x ~ e^{i \gamma x^3/3} | x \rangle. 
\end{eqnarray}
A direct implementation of the cubic phase gate requires strong nonlinearity, such as in a nonlinear crystal, which is very challenging 
for current experimental techniques. Another way to implement the cubic phase gate is based on the adaptive non-Gaussian measurement (AnGM) \cite{angm} method 
with a resource state, e.g., the cubic phase state. This method either requires Gaussian feedforward or post selection, the latter implies that it is nondeterministic. However, this method can be well fitted into the measurement-based quantum computation, which itself is based on teleportation. 

\begin{figure}[ht!]
\includegraphics[width=\columnwidth]{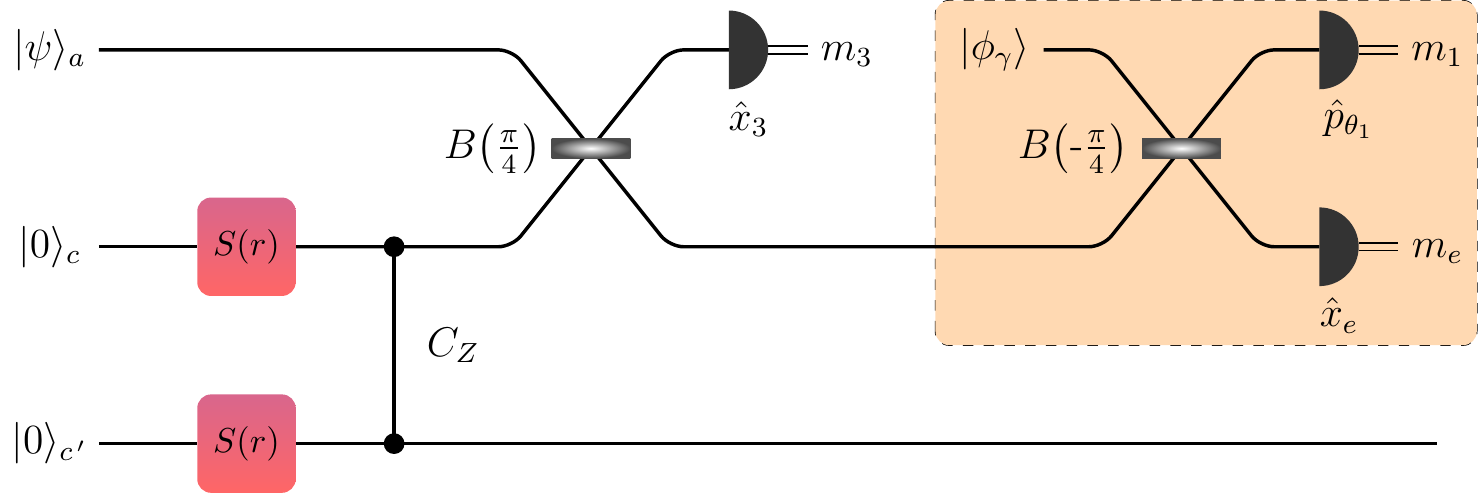}
\caption{ Teleportation circuit that implements a cubic phase gate \cite{Alexander2018Universal}. 
A cubic phase circuit (the part in the orange shaded box) consists of a beam splitter
$B \big({\rm -}\frac{\pi}{4} \big)$ (a 50:50 beam splitter with an additional $\pi$ phase shift), two homodyne detectors that measure quadratures $\hat x_e$ and $\hat p_{\theta_1}$, and an input cubic phase state $| \phi_{\gamma} \rangle$. The whole 
teleportation circuit is a generalisation of the gate teleportation circuit Fig. \ref{fig:gate-teleportation}, which implements a single-mode Gaussian unitary, by replacing 
one of the homodyne detectors by the cubic phase circuit. Note that another homodyne detector measures the position quadrature $\hat x_3$. 
} 
\label{fig:cubic-phase-single}
\end{figure}

While there are many types of teleportation circuits, we focus on a particular one which is shown in Fig. \ref{fig:cubic-phase-single} \cite{angm,Alexander2018Universal}. 
This is basically a generalisation of the circuit in Fig. \ref{fig:gate-teleportation} that implements measurement-based Gaussian unitaries. The ideal cubic phase state 
$| \phi_{\gamma} \rangle$ is used as a resource state to implement the cubic phase gate. In the infinite squeezing limit, the circuit in Fig. \ref{fig:cubic-phase-single} implements
a unitary \cite{Alexander2018Universal}
\begin{eqnarray}\label{eq:cubic-phase-gate}
\hat L (\gamma, \sigma, \boldsymbol m) = Z \big(\sqrt{2} m_3 \big) X (\kappa) R \bigg( -\frac{\pi}{2}\bigg) P (\tau) V \big( -2 \sqrt{2} \gamma \big), \nonumber\\
\end{eqnarray}
where 
\begin{eqnarray}
\tau &=& 4 \sigma + 4\gamma \big(m_3 + \sqrt{2} m_e \big), \nonumber\\
\kappa &=& -2m_1 \sqrt{1+\sigma^2} - 2\sigma \big(\sqrt{2}m_3 + m_e \big) \nonumber\\
&& - \sqrt{2} \gamma \big( m_3 + \sqrt{2} m_e \big)^2, \nonumber\\
\sigma &=& \tan \theta_1, 
\end{eqnarray}
and the overall phase has been neglected. Here $ P(\tau) = \exp \big(i \tau \hat x^2/2 \big)$ is known as the shear operator. From Eq. \eqref{eq:cubic-phase-gate}
it is evident that a cubic phase gate is implemented, as well as a series of Gaussian unitaries: displacements, phase shift and local shear. To implement only a cubic phase gate,
Gaussian feedforward corrections need to be applied after the homodyne detection. \\

\noindent \textit{Simplification using adaptivity.} At every step of the computation and hence the measurement outcome, there are measurement dependent displacements that are produced. These can be accumulated and corrected at the end of the computation. With the cubic phase gate, one can adjust the angle of the homodyne measurement to account for any quadratic phase gates that results from commuting the displacements across the cubic gate implementation operator as in Eq. \eqref{eq:cubic-phase-gate}, which was already shown in Ref. \cite{Alexander2018Universal}. 

\subsubsection{Resource states for approximate/weak cubic phase gate}\label{Sec:WeakCubicPhase}
There are various resource states that have been proposed in the literature (for a review see Ref. \cite{Krishna2018ON}). We briefly go through a few of the proposals that are suited for implementation in the cluster architecture. Note from Eq. \eqref{e4} \cite{Alexander2018Universal} that the effective non-Gaussian operator $ \phi_r(\sqrt{2}m_e-\hat{x})$ is in principle similar to the implementation of a GKP circuit \cite{Krishna2018ON,gkp}, which implements the transformation $\phi_r(\hat{x}+q)$, where $q$ is the output of a homodyne measurement in the GKP circuit. The advantage of the AnMG \cite{angm}, when compared to the GKP method, is that the quadratic feedforward corrections that are required can be incorporated into the measurement, whereas for the GKP method. It would require a dynamic quadratic phase gate as a feedforward correction. Both methods require a measurement dependent displacement operation. \\

\noindent \textbf{ON states.}
We briefly introduce the use of ON states \cite{Krishna2018ON} as a resource for implementing weak quadrature phase gates given by $\exp\big(i \gamma \hat x^N\big)$, where
$N$ is the order of the gate. A general ON state is defined as 
\begin{align}
\ket{ON} = (1+|a|^2)^{-1/2} \left(\ket{0} + a \ket{N} \right), ~a \in \mathbb{C}\label{eqn:ON},
\end{align}
where the Fock excitation $N$ is also the order of the gate we want to implement. 
With respect to the implementation of a cubic phase gate we consider as resource an ON state with $N=3$. The $\ket{03}$ state is defined as $\ket{\phi_r} = c_a (\ket{0} + a \ket{3}), \, c_a = (1 + |a|^2)^{-1/2}$. The ON states also serve as a transparent example of how the non-Gaussianity of the resource states in its wave-function gets transferred to the corresponding non-Gaussian operator that is applied to the respective node of the cluster. Choosing $a = -i\gamma \sqrt{3}/2$, $|\gamma| \ll 1$, and following the steps mentioned in Ref. \cite{Krishna2018ON} we have that 
\begin{align}
\phi_r(\kappa-\hat{x}) = \hat{A}_{\kappa} Z(3\gamma (\kappa^2-1/2) ) P(-3\gamma \kappa) V(\gamma),
\label{e9}
\end{align}
where $\hat{A}_{\kappa} = \exp [-(\kappa -\hat{x})^2/2] $ is an unavoidable measurement dependent Gaussian noise factor that results from the choice of resource state (and can be interpreted as a type of finite squeezing effect). Note that we have also neglected the phase factors and overall scaling constants in Eq. \eqref{e9}. We can then substitute the above expression into Eq. \eqref{e4} to get the final effective operation of using the $\ket{ON} =\ket{03}$ state as the ancilla non-Gaussian state in the measurement. The Gaussian elements can all be commuted through the non-Gaussian operation to the left to be corrected at the end of the commutation. We can implement stronger cubic gates by repeating this procedure. This method in principle can be used for higher-order gates using suitable ON states with higher values of $N$. Also a generalised version to implement higher-order non-linear gates using the measurement based computation has been recently proposed in Ref. \cite{h-order}. \\

\noindent \textbf{GKP state.} The GKP state was the first proposal for the approximate preparation of the cubic phase state \cite{gkp}. Here, one arm of a two mode squeezed state is displaced along the momentum quadrature and then measured using a photon number resolving detector. Then depending on the measurement outcome, the other arm is squeezed to generate the cubic phase state of interest, as shown in Eq. \eqref{eq:GKPstate} below:
\begin{align}\label{eq:GKPstate}
 \mbox{
\Qcircuit @C=0.5em @R=2em { 
\lstick{\ket{0}}& \qw&\gate{S}& \multigate{1}{B(\pi/4)} & \qw & \gate{Z(w)}&\measureD{\mbox{$\Pi_{n}$}} & \ustick{m}  \controlo \cw  \cwx[1] \\ 
\lstick{\ket{0}}& \qw&\gate{S^{-1}}&\ghost{B(\pi/4)} & \qw & \qw&\qw &\gate{S(m)} & \rstick{\ket{\phi_r} ~.} \qw  }}
\end{align}
This is a possible candidate for use as an ancilla resource, as shown in Fig.~\ref{fig:cubic-phase-single}. 
It turns out that the approximation works well only in the limit of large initial squeezing, displacement and photon-measurement outcome. A later analysis of this procedure put forth some of the experimental challenges of this method \cite{ghose-sanders}. 

Since the resource state directly approximates the cubic phase state, its use in the generalised measurement in the cluster will implement an operation that approximates Eq. \eqref{eq:cubic-phase-gate}. \\

\noindent \textbf{MFF state.} The  resource state proposed by Marek-Filip-Furasawa (MFF) in Refs. \cite{angm, mff1} is of the form $\ket{\phi_r} = {S}\big(1+i\gamma \hat{x}^3 \big)\ket{0}$, where 
$\big(1+i\gamma \hat{x}^3 \big)$ is the first-order Taylor expansion of a cubic phase gate $\exp \big(i \gamma \hat{x}^3 \big)$ and $S$ is the squeezing operator. 
To implement higher-order expansions one has to then generate the appropriate resource state by applying the suitable Taylor expansion operator on vacuum. \\

\noindent \textbf{Gaussian optimised state} \cite{angm}. Here, the authors proposed a state which is a superposition of the first four Fock basis states after performing a Gaussian operation on their test states, i.e., a state of the form 
\begin{align}
\ket{\phi_r} = U_G \left[ \sum_{i=0}^3 c_i \ket{i}\right].
\end{align}
This use of Gaussian optimised non-Gaussian state preparation was earlier proposed in Ref. \cite{gaussopt}. The reason that the superposition was truncated at $n=3$ is because it is possible to experimentally prepare these states in the lab \cite{akira-upto3}. The optimisation was performed by minimising the cost function which was the variance of the non-linear quadrature (NLQ) operator \cite{angm}
\begin{align}
\hat{p}_{\rm NLQ} = \hat{p}^2 - 3 \gamma \hat{x}^2,
\end{align}
and the mean $\langle\hat{p}_{\rm NLQ}\rangle$. 

\subsubsection{Additional methods for non-Gaussian operations in the cluster}
Non-Gaussianity is an important resource to have in the cluster state with regard to universal quantum computation \cite{Walschaers2018nonGaussianCluster, Ohliger2010limitation}.
We now present two ways to inject non-Gaussianity into the cluster. The first method we explore is what we deem the cluster-gate model, where an off-cluster gate model is used to implement non-Gaussian operations on particular cluster modes in case it is advantageous over the measurement-based method discussed earlier. The second method is to use the already existing optical elements in the cluster architecture but switching the homodyne detectors for photon number-resolving detectors. The final possible location non-Gaussian resources could be incorporated to generate a non-Gaussian cluster state is at the level of the source. We however explore this avenue elsewhere. This subsection entirely focuses on the one-dimensional cluster but the methods can be suitably generalised to the two-dimensional cluster where one can `double' the effects by repeating these methods on the remaining modes in the cluster generation circuit. \\

\begin{figure}[ht!]
\includegraphics[width=\columnwidth]{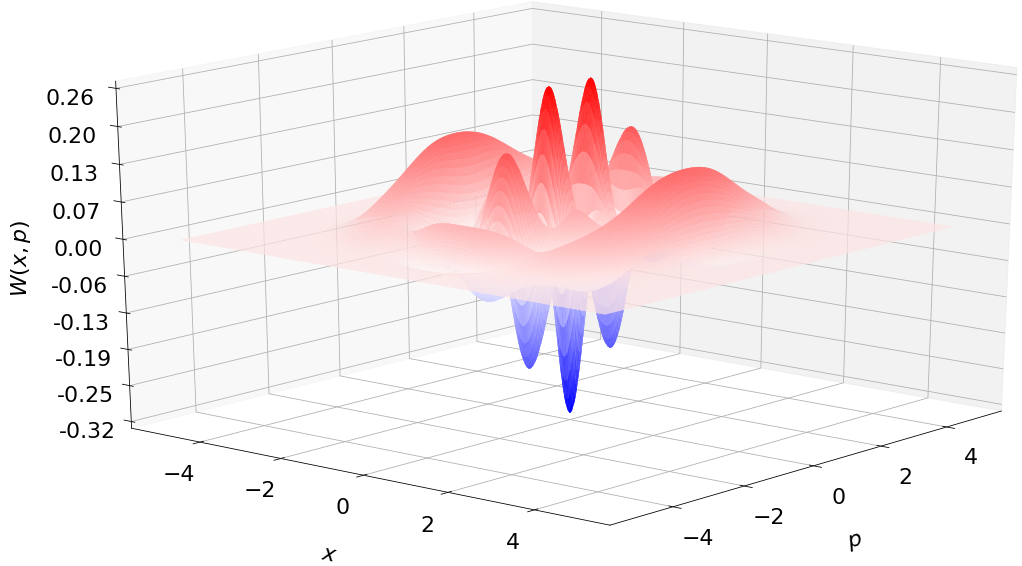}
\caption{Wigner function for an input displaced squeezed state with $(\alpha,\xi) = (0.02+i\,0.84,0.5)$ and the two-mode squeezed state with squeeze parameter $r=0.68$, and the PNR measurements $(\kappa_1,\kappa_2)$ reading $(3,2)$, respectively.  The Wigner function was computed using the Strawberry Fields software \cite{SFpaper}.
 } 
\label{fig:pnrsq}
\end{figure}

\begin{figure}[ht!]
\includegraphics[width=\columnwidth]{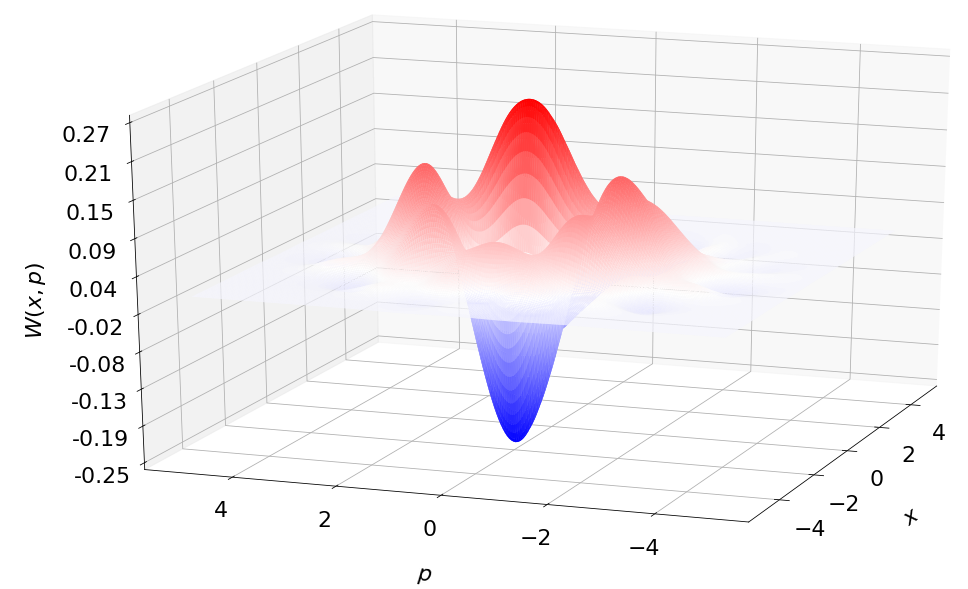}
\caption{Wigner function for a coherent state with $\alpha = 0.02+i\,0.84$ and the squeezing parameter of the two-mode squeezed state given by $r=0.68$, with PNR measurements $(\kappa_1,\kappa_2)$ reading $(4,1)$, respectively. The Wigner function was computed using the Strawberry Fields software \cite{SFpaper}.
} 
\label{fig:pnrcoh}
\end{figure}

\noindent \textbf{Hybrid cluster-gate model.} It is possible that certain realistic implementations of the cubic phase gate may be noisy or challenging to implement using the ancilla-assisted measurement-based model. One could in principle use a switch to take out a particular mode and perform a gate directly and feed it back into the cluster as shown in the circuit  below: 
\begin{align}
 \mbox{
\Qcircuit @C=0.7em @R=2.0em { 
\lstick{{\rm switch~out} ~~\ket{\psi}} & \qw&\multigate{1}{~U~} & \qw && ~~~~~~ T\ket{\psi}~{\rm to~switch~in}\\
\lstick{\ket{\phi_r}}& \qw&\ghost{~U~} & \qw &\measureD{\mbox{$\Pi$}} & \cw ~~~~~~~~~m ~~~~.
}}
\end{align}
 With this scheme, one could in principle apply several variations of the implementation of the cubic phase gate mentioned in Table II of Ref. \cite{Krishna2018ON}. In the event the gate model produces better quality/fidelity of the cubic phase gate, this hybrid approach could prove advantageous. 

\begin{figure*}[ht!]
\includegraphics[width=14cm]{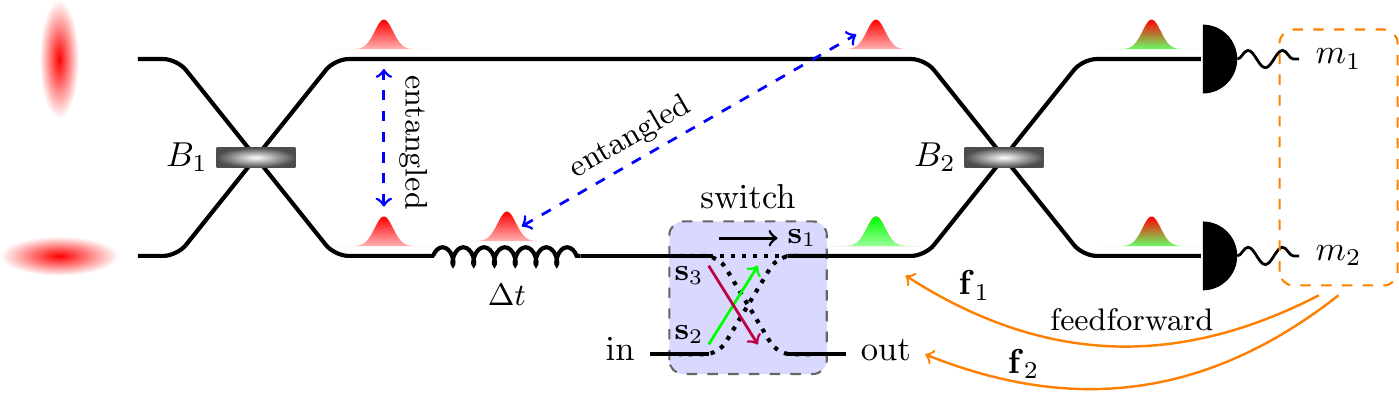}
\caption{ Optical setup that implements gate teleportation, or a series of single-mode unitaries \cite{Alexander2017MBLO}. A switch has been added as compared to Fig. \ref{fig:1D-cluster-setup}, which 
allows for injecting input states and reading out output states. The states of the switch ({\bf s$_1$}, {\bf s$_2$} and {\bf s$_3$}) have to be set accordingly during the computational process. 
The homodyne measurement outcomes ${m}_1$ and ${m}_2$ are used to do feedforward. There are two ways of doing feedforward: {\bf f$_1$} and {\bf f$_2$}. {\bf f$_1$} means doing feedforward
after every measurement step, while {\bf f$_2$} means doing feedforward at the end of computation. If only Gaussian unitaries are implemented, both ways are 
allowable. However, if non-Gaussian unitaries are implemented one has to do {\bf f$_1$}. } 
\label{fig:1D-cluster-computation-setup}
\end{figure*}

In the ancilla-based computation, the gate is applied by using a switch at one of the homodyne detectors and applying a general adaptive non-Gaussian measurement that implements a non-Gaussian operation on a cluster node. In the proposed hybrid model, we use the switch of Fig. \ref{fig:1D-cluster-computation-setup} first in mode {\bf s$_3$}, then we apply the gate and feed it into the cluster using mode {\bf s$_2$} of the switch. Then one can use the cluster and perform the measurement-based computation on the modes. \\

The current state-of-the-art generation of squeezed pulses is at a repetition rate in the GHz range \cite{Mangold2014}. Other experimental elements that would have an effect on the rate at which computation can be performed include optical switches, adaptive homodyne measurements, and non-Gaussian resource generation elements such as photon number resolving detectors. \\

\noindent \textbf{PNR induced non-Gaussian operations.} While it is necessary to have precise control over the gates for specific algorithms where the gate order is crucial, it turns out that having any non-linearity in the cluster would still be advantageous for other applications such as quantum machine learning \cite{Biamonte2017}. To this end, we obtain the operator that is implemented when the homodyne measurements are replaced by PNR detectors using a switch between the two types of measurements.

\begin{figure}[ht!]
\includegraphics[width=\columnwidth]{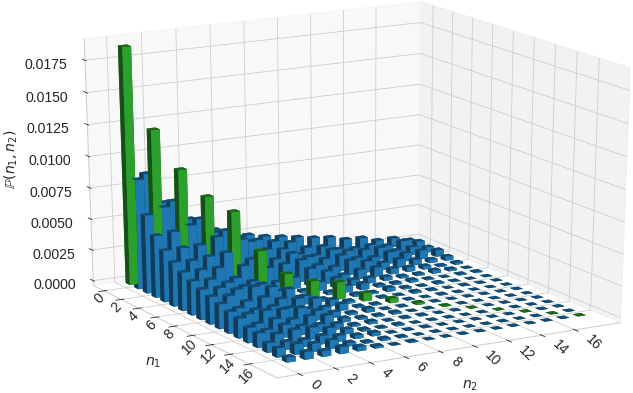}
\caption{Measurement probabilities $P(n_1,n_2)$ for input displaced squeezed state with $(\alpha,\xi) = (0.02+i\,0.84,0.5)$ and the two-mode squeezed state with squeeze parameter $r=0.68$. Measurement outcomes with $\kappa_1=\kappa_2$ are more favorable. 
The probability was computed using the Strawberry Fields software \cite{SFpaper}.
} 
\label{fig:pnrsqprob}
\end{figure}

\begin{figure}[ht!]
\includegraphics[width=\columnwidth]{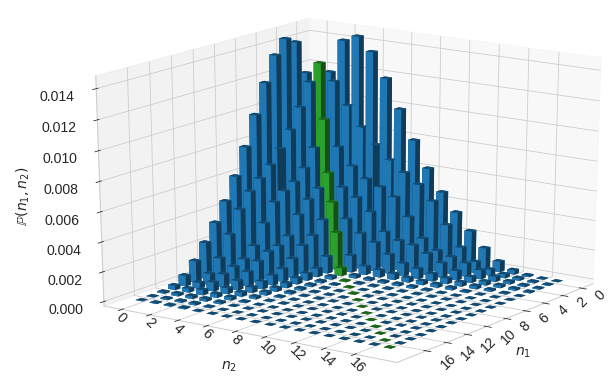}
\caption{Measurement probabilities $P(n_1,n_2)$ for an input coherent state with $\alpha = 0.02+i\,0.84$ and the squeezing parameter of the two-mode squeezed state given by $r=0.68$. Measurement outcomes are observed to be skewed away from $\kappa_1 = \kappa_2$.
The probability was computed using the Strawberry Fields software \cite{SFpaper}.
} 
\label{fig:pnrcohprob}
\end{figure}

The basic one-dimensional cluster circuit with the homodyne measurements replaced by the photon-number resolving (PNR) detectors is given by 
\begin{align} \label{psub} 
 \mbox{
\Qcircuit @C=0.45em @R=1.5em { 
\lstick{\ket{\psi_{\rm in}}} & \qw &\qw&\qw&\multigate{1}{B(\pi/4)}&  \measureD{\Pi_n} & \cw ~~~~ \kappa_1\\ 
\lstick{\ket{0}} &\gate{S(r)}&\multigate{1}{B(\pi/4)} &\qw& \ghost{B(\pi/4)}&\measureD{\Pi_n}& \cw ~~~~ \kappa_2\\
\lstick{\ket{0}}&\gate{S(r)^{-1}}& \ghost{B(\pi/4)}& \qw&\qw&\qw & ~~~~\widehat{T}_{\kappa_1,\kappa_2} \ket{\psi_{\rm in}}.\\
}}
\end{align}
The single-mode squeezers $S(r)$ and the $50:50$ beam splitter $B(\pi/4)$ action on the two vacuum states produces the standard two-mode squeezed state $\ket{\psi_{\rm sq}(r)}$. So we need to evaluate 
\begin{align}
&\widehat{T}_{\kappa_1,\kappa_2} \ket{\psi_{\rm in}} =  [\bra{\kappa_1} \bra{\kappa_2} ][ B(\pi/4) \otimes \mathbb{I}][\ket{\psi_{\rm in}} \ket{\psi_{\rm sq}(r)}], \nonumber \\
& \ket{\psi_{\rm sq}(r)} = {\rm sech\,r} \sum_{j=0}^{\infty} ({\rm tanh\,r})^n \ket{jj}.
\label{pnrops}
\end{align}
The matrix elements of the beam splitter were previously derived in \cite{kraus10} and employed in \cite{nong} to study Gaussian and non-Gaussian channels. We have that 
\begin{align}
B(\pi/4) = \sum_{m_1=0}^{\infty} \sum_{m_2 =0}^{\infty} \sum_{n_1 =0}^{\infty} \sum_{n_2 =0}^{\infty} C^{m_1m_2}_{n_1n_2} \ket{m_1 m_2} \bra{n_1n_2},
\end{align}
where 
\begin{align}
C^{m_1m_2}_{n_1n_2} &= \sqrt{\frac{m_1!m_2!}{n_1!n_2!}} \sum_{r=0}^{n_1} \sum_{j=0}^{n_2} \binom{n_1}{r} \binom{n_2}{j} (-1)^{n_2-j} \nonumber\\
& \times 2^{-(n_1+n_2)} \delta_{m_2,r+j} \,\,\delta_{m_1,n_1+n_2-r-j}.
\label{coeff}
\end{align}
Substituting Eq. \eqref{coeff} into Eq. \eqref{pnrops} and after some simplification we get that 
\begin{align}
\ket{\psi_{\rm out}} &= \widehat{T}_{\kappa_1,\kappa_2} \ket{\psi_{\rm in}} \propto  \sum_{n=0}^{\kappa_1+\kappa_2} \sum_{r=r_{\rm min}}^{r_{\rm max}} \sum_{j=0}^n ({\rm tanh\,r})^n (-1)^{n-j} \nonumber\\
& \times \binom{\kappa_1+\kappa_2 -n}{r} \binom{n}{\kappa_2-r} \braket{\kappa_1+\kappa_2 -n}{\psi_{\rm in}} \ket{n},
\end{align}
where 
\begin{align}
r_{\rm min} &= \max\{0,\kappa_2-n\}, \, \nonumber \\
r_{\rm max} &= \max\{\kappa_2,\kappa_1+\kappa_2 -n\}.
\end{align}
The final physical state is then given by normalising $\ket{\psi_{\rm out}}$. 

We now provide some examples of the resulting action of the PNR measurements on different input modes. The Wigner function of the output for measurement outcomes $(\kappa_1,\kappa_2) = (3,2)$ for an input coherent state is given in Fig. \ref{fig:pnrsq} and outcomes $(4,1)$ for a squeezed state in Fig. \ref{fig:pnrcoh}.

In Fig. \ref{fig:pnrsqprob} we also plot the probability of obtaining the various Fock outcomes for the input squeezed state. We observe that for a fixed total number of photon detections, the measurement outcomes  $\kappa_1 = \kappa_2$ are more favorable. Similarly, we plot the probability distribution for various PNR measurement outcome pairs for the input coherent state in Fig.\,\ref{fig:pnrcohprob}. We find that the measurements tend to be skewed away from $\kappa_1= \kappa_2$.

\subsection{Teleportation along a one-dimensional temporal CV cluster state}
The main purpose of generating CV temporal cluster states is to achieve quantum computation, e.g., to implement unitary transformations. To do that we have to inject 
input states into the cluster  and readout the output states after performing some unitary transformations. The one-dimensional CV temporal cluster states are not 
universal because they can only be used to implement single-mode unitaries. However, it is instructive to introduce how to inject input states, implement single-mode
unitaries via homodyne measurements and readout output states based on one-dimensional cluster states. 
The generalisation to universal temporal CV cluster states (at least two-dimensional cluster) is then straightforward. 

The strategy is to add a switch after the delay loop, as shown in Fig. \ref{fig:1D-cluster-computation-setup}. The switch has three states, denoted as {\bf s$_1$}, {\bf s$_2$} and {\bf s$_3$}. 
At state {\bf s$_1$}, the delay loop is connected to the beam splitter $B_2$, and the input and output wires are disconnected to the optical setup. At state {\bf s$_2$}, the input wire is 
connected to the beam splitter $B_2$ and input states can be injected into the optical setup. At state {\bf s$_3$}, the output wire is connected to the delay loop and optical fields
from the delay loop can be readout or detected. The states of the switch are set accordingly during the process of computation. 

Measurement-based quantum computation is based on the gate teleportation, which has been discussed in Sec. \ref{sec:gate-teleportation}. 
The gate teleportation protocol described in Fig. \ref{fig:gate-teleportation} can be implemented in the optical setup in Fig. \ref{fig:1D-cluster-computation-setup}. 
This is achieved by injecting only one pair of single-mode squeezed pulses and setting the switch in the state {\bf s$_2$} to the let the input mode couple with the top 
mode. After performing a feedforward to the delayed mode using the homodyne measurement outcomes, the state of the delayed mode is transformed into 
the input state applied by a particular unitary. 

\begin{figure}[htp]
\subfloat[]{%
  \includegraphics[clip,width=1\columnwidth]{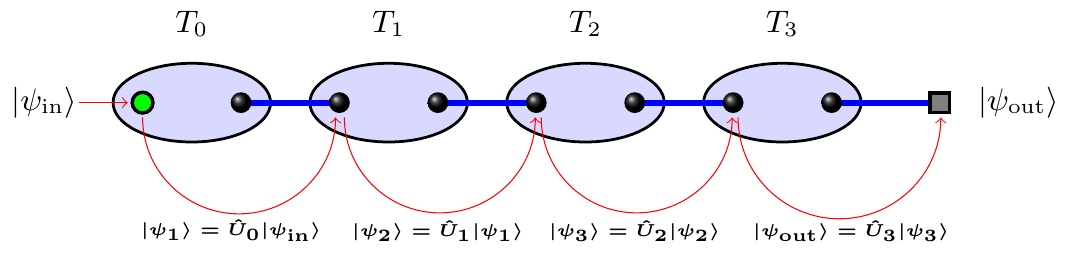}%
}

\subfloat[]{%
  \includegraphics[clip,width=1\columnwidth]{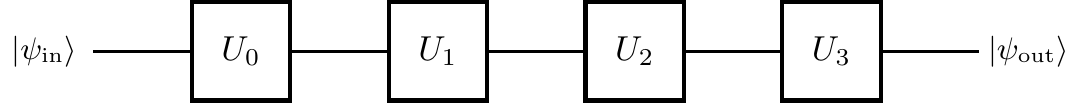}%
}

\caption{Implementation of four single-mode unitaries using one-dimensional temporal cluster states \cite{Alexander2017MBLO}. (a) Four single-mode unitaries $\hat U_0$, 
$\hat U_1$, $\hat U_2$ and $\hat U_3$ are implemented sequentially via four steps of homodyne measurements. $T_i$ ($i = 0, 1, 2, 3$) is the time of performing the 
beam splitter transformation and homodyne measurements, and is also used as the notation of the macronode. We assume that $T_i < T_{i+1}$, which means the time
direction is from left to right and is different from that shown in Fig. \ref{fig:1D-cluster-state}. (b) The corresponding gate model circuit.  }
\label{fig:four-unitary-teleportation}
\end{figure}

A sequence of gate teleportations, thus a sequence of single-mode unitaries, can be implemented by injecting a series of pairs of single-mode squeezed states and appropriately 
setting the states of the switch. Fig. \ref{fig:four-unitary-teleportation} shows an example of implementing four single-mode uintaries, and Table \ref{tab:teleportation-four-unitary} 
lists the states of the switch and relevant actions during the computation process. If only Gaussian unitaries are implemented, the feedforward can be done at the end. We next move on to the generation of two-dimensional cluster states and performing computation using them. 


\begin{table*}[tp]
\caption{Process of implementing four single-mode uintaries.} 
\label{tab:teleportation-four-unitary}\centering%
\begin{center}
    \begin{tabular}{| c | c | l |}
    \hline
    Time (macronode) & ~State of switch~ & Description \\ \hline 
    $T_0$ & {\bf s$_2$} & Inject input state/Homodyne measurements/Feedforward to delayed mode \\ \hline 
    $T_1$ & {\bf s$_1$} & Homodyne measurements/Feedforward to delayed mode \\ \hline 
    $T_2$ & {\bf s$_1$} & Homodyne measurements/Feedforward to delayed mode \\ \hline 
    $T_3$ & {\bf s$_3$} & Homodyne measurements/Feedforward to output mode/Readout output state \\ 
    \hline
    \end{tabular}
\end{center}
\end{table*}

\section{Two-dimensional temporal cluster states}\label{sec:2Dcluster}

\subsection{Generation of two-dimensional temporal cluster state}

\begin{figure*}[ht!]
\includegraphics[width=16cm]{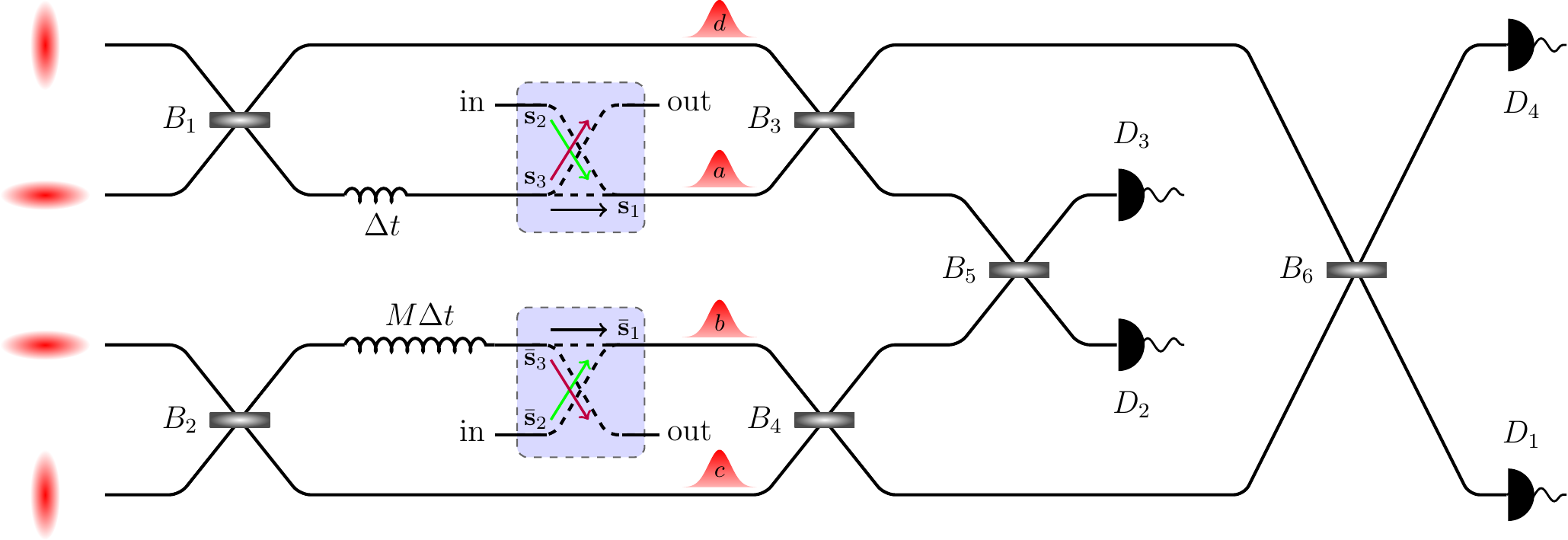}
\caption{ Optical setup that generates two-dimensional temporal cluster states and implements measurement-based quantum computation \cite{Alexander2017MBLO}. 
The whole setup consists of 
six $50:50$ beam splitters $B_i (i=1, \cdots, 6)$, four homodyne detectors $D_k (k=1, 2, 3, 4)$, two delay loops (one with delay $\Delta t$ and the other $M\Delta t$ where
$M$ is the depth of the circuit) and 
two switches, each of which has three states.  The notation of the states of the top switch is the same as that introduced in Fig. \ref{fig:1D-cluster-computation-setup}, while 
the states of the bottom switch are denoted as $\bar {\boldsymbol{\rm s}}_1$, $\bar {\boldsymbol{\rm s}}_2$ and $\bar {\boldsymbol{\rm s}}_3$. When generating cluster states, the switches are set to $\boldsymbol{\rm s}_1$
and $\bar {\boldsymbol{\rm s}}_1$. A series of four squeezed pulses with appropriate phases are injected continuously into the setup. When implementing measurement-based unitaries,
the states of the switches have to be set accordingly to allow injecting input states and reading out output states, and homodyne measurement outcomes have to be used
to do feedforward. Four modes before entering the beam splitters $B_3$ and $B_4$ are depicted and denoted as modes $d, a, b, c$ (from top to bottom). } 
\label{fig:2D-cluster-setup}
\end{figure*}

Two-dimensional temporal cluster states can be generated by injecting four squeezed pulses into the optical setup shown in 
Fig. \ref{fig:2D-cluster-setup}. The first (top) pair of 
single-mode squeezed pulses are injected into an optical circuit (leading to the beam splitter $B_3$) that is the same as 
Fig. \ref{fig:1D-cluster-setup} and a one-dimensional temporal cluster 
state is generated. The second (bottom) pair of  single-mode squeezed pulses are injected into an optical part (leading to the beam splitter $B_4$) that is also the same as Fig. \ref{fig:1D-cluster-setup} but with a delay $M\Delta t$, with $M$ a positive integer.  This gives rise to another one-dimensional 
temporal cluster state. Finally, the beam splitters $B_5$ and $B_6$ lead to
the second dimension of a two-dimensional temporal cluster state. The depth of the two-dimensional 
cluster state depends on $M$, namely, the length of the second delay loop.  

\begin{figure}[ht!]
\includegraphics[width=8.3cm]{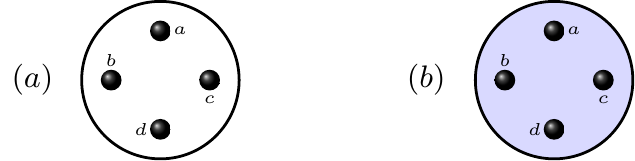}
\caption{ Additional graph elements. (a) The black dots represent optical modes as before and the labels show their correspondence in Fig. \ref{fig:2D-cluster-setup}. The
black circle represents the action of four beam splitters $B_3$, $B_4$, $B_5$ and $B_6$. (b) The black circle filled with light blue represents the action of the four beam splitters
followed by homodyne measurements. } 
\label{fig:4mode-macronode}
\end{figure}

\begin{figure*}[ht!]
\includegraphics[width=10cm]{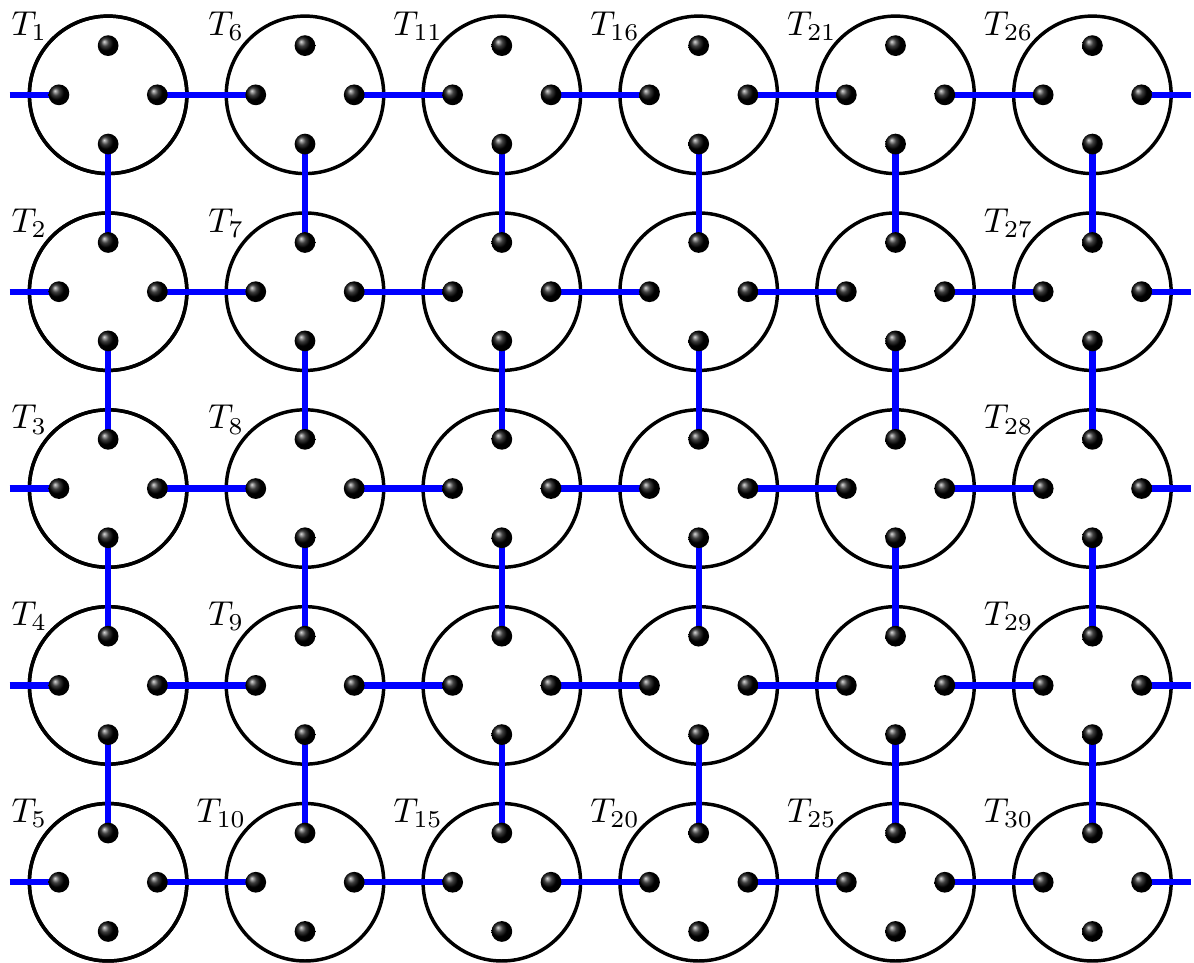}
\caption{ Graphical representation of a two-dimensional temporal cluster state \cite{Alexander2017MBLO} with depth five ($M=5$). $T_i$ denotes the time to perform the four-beam-splitter transformation and it 
also acts as the label of the corresponding macronode. It is assumed that $T_i < T_{i+1}$ and $T_{i+1} - T_i = \Delta t$. Note that the last mode of each column should be 
connected to the first mode of the next column, which is not plotted in the figure for convenience. } 
\label{fig:2D-cluster-state-graph}
\end{figure*}

A complete graphical representation of the two-dimensional temporal cluster state was developed in Ref. \cite{Menicucci2011}. A simplified version was discussed in Ref. \cite{Alexander2016}
and we will use the simplified version in this paper. To discuss the graphical representation, we first introduce additional graph elements, 
as shown in Fig. \ref{fig:4mode-macronode}.  As before, the black dots represents the optical modes, in particular, the optical modes before entering the beam splitters 
$B_3$ and $B_4$, as shown in Fig. \ref{fig:2D-cluster-setup}. The black circle represents the action of the four beam splitters $B_3$, $B_4$, $B_5$ and $B_6$. If the black circle
is filled with colours (light blue in Fig. \ref{fig:4mode-macronode}), it means homodyne measurements are performed to the four modes. The black circle and the four modes it encircles are defined as a macronode. 
With these graph elements in hand, together with those introduced in Fig. \ref{fig:graph-elements}, we can draw a graph for any two-dimensional temporal cluster state. 
Fig. \ref{fig:2D-cluster-state-graph} shows an example with $M=5$. 

\subsection{Measurement-based two-mode Gaussian unitary}\label{sec:MBGU}

\begin{figure}[ht!]
\includegraphics[width=8.5cm]{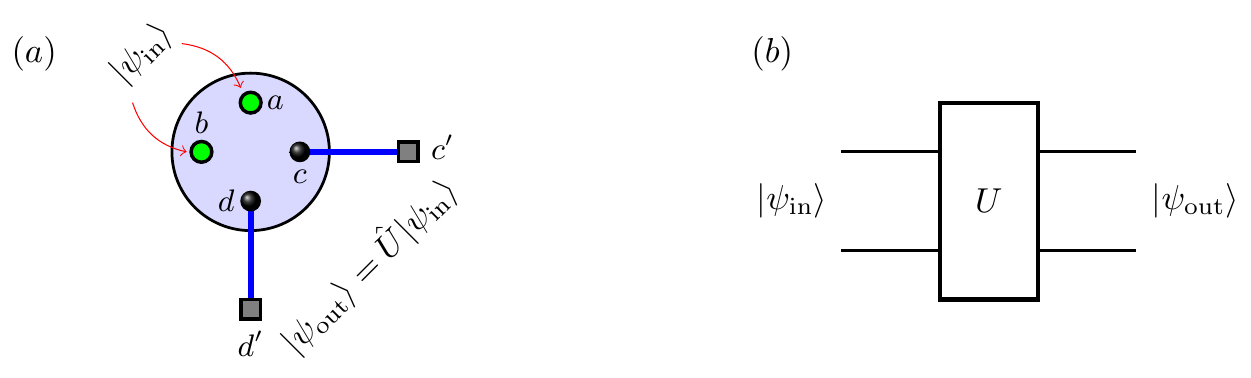}
\caption{ Two-mode Gaussian unitaries via  four-mode homodyne measurements  \cite{Alexander2017MBLO} . 
(a) Two input modes (modes $a$ and $b$) and two optical modes $c$ and $d$ couple via 
four beam splitters, and are detected by homodyne detectors. Depending on the measurement angles of the homodyne detectors, a two-mode Gaussian unitary is implemented. 
The output state comes out via the output modes $c'$ and $d'$. (b) Gate model representation of the corresponding two-mode Gaussian unitary. } 
\label{fig:teleportation-graph-2mode}
\end{figure}

Four-mode homodyne detection, as shown in Fig. \ref{fig:2D-cluster-setup}, can implement an arbitrary two-mode Gaussian unitary by appropriately choosing the measurement
quadratures (angles). A detailed discussion was given in Ref. \cite{Alexander2016}. Fig. \ref{fig:teleportation-graph-2mode} shows the graphical representation of the four-mode homodyne 
measurement. In the limit of infinite squeezing in the source squeezed pulses, the two-mode unitaries can be implemented perfectly. However, for physical squeezed pulses, the amount
of squeezing is finite. This results in errors when implementing unitaries via homodyne measurements. If the input states are known, the error due to the effect of finite squeezing can be corrected \cite{Su2018EC}. For simplicity, we first ignore the finite-squeezing effect and work in the infinite squeezing limit.  

The four homodyne measurements implement a Gaussian unitary $\hat G_{jk}$ \cite{Alexander2016}, 
\begin{eqnarray}
\hat G_{jk} (\boldsymbol m, \boldsymbol \theta) &=& \hat B_{jk}^{\dag}(\pi/4) \hat A_j (m_1, m_3, \theta_1, \theta_3) \nonumber\\ 
&&\times \hat A_k (m_2, m_4, \theta_2, \theta_4) \hat B_{jk}(\pi/4),
\end{eqnarray}
where $\boldsymbol m = (m_1, m_2, m_3, m_4)$, $\boldsymbol \theta = (\theta_1, \theta_2, \theta_3, \theta_4)$ and 
\begin{eqnarray}
\hat A_j (m_h, m_l, \theta_h, \theta_l) = D_j (m_h, m_l, \theta_h, \theta_l) \hat U_j (\theta_h, \theta_l). 
\end{eqnarray}
$D_j (m_h, m_l, \theta_h, \theta_l)$ is the phase-space displacement operator
\begin{eqnarray}
 D_j (m_h, m_l, \theta_h, \theta_l) = \hat D \bigg[ \frac{-i e^{i\theta_l} m_h - i e^{i \theta_h} m_l}{\sin \big( \theta_h - \theta_l \big)} \bigg]
\end{eqnarray}
and $\hat U_j (\theta_h, \theta_l)$ is a single-mode unitary
\begin{align}
\hat U_j (\theta_h, \theta_l) &= R_j(\theta_{h,l}^+) S_j \big[ {\rm ln} \big(\tan \theta_{h,l}^- \big) \big] R_j(\theta_{h,l}^+),
\end{align}
where $\theta_{h,l}^{\pm} = \big(\theta_h \pm \theta_l\big)/2$.
The displacements can be corrected by applying feedforward corrections conditioned on the homodyne measurement outcomes, so the implemented two-mode Gaussian unitary is 
\begin{eqnarray}\label{eq:MB-2mode-unitary}
\hat U_{jk} (\boldsymbol \theta) = \hat B_{jk}^{\dag}(\pi/4) \hat U_j (\theta_1, \theta_3) \hat U_k (\theta_2, \theta_4) \hat B_{jk}(\pi/4),
\end{eqnarray}
which is completely determined by the homodyne measurement angles. 

In the case that $\theta_1 = \theta_2$ and $\theta_3 = \theta_4$, the two same single-mode unitaries commute with the beam splitter operator such that a pair of 
same single-mode unitaries is implemented, namely,
\begin{eqnarray}
\hat U_{jk} (\theta_1, \theta_3) = \hat U_j (\theta_1, \theta_3) \hat U_k (\theta_1, \theta_3). 
\end{eqnarray}

\noindent {\bf Phase shifts}: If further $\theta_1 = \theta_3 = \theta$, then a pair of same phase shifts is implemented, $\hat U_{jk} (\theta) = R_j(2\theta) R_k(2\theta)$. \\

\noindent {\bf Squeezers}: If further $\theta_1 = -\theta_3 = \theta$, then a pair of same single-mode squeezers is implemented, 
$\hat U_{jk} (\theta) = S_j \big[ {\rm ln} (\tan \theta) \big] S_k \big[ {\rm ln} (\tan \theta) \big]$.\\

To get rid of the beam splitters in Eq. \eqref{eq:MB-2mode-unitary}, the implemented single-mode unitaries in two modes have to be the same. This constraint is less desirable because 
generally we want to implement different single-mode unitaries in different modes. This difficulty can be overcome by sequentially implementing the two-mode unitary Eq. \eqref{eq:MB-2mode-unitary} such that the beam splitters between two neighbouring unitaries cancel. The resulting two-mode unitary is a sequence of different single-mode unitaries in 
each mode sandwiched by two beam splitters. The two beam splitters can be further cancelled by another two steps of homodyne measurement (one before and the other after the
sequence of two-mode unitaries) because the homodyne measurements can also induce a beam splitter transformation. 

In the case of $\theta_3 = \theta_1 - \pi/2$ and $\theta_4 = \theta_2 - \pi/2$, the two-mode unitary Eq. \eqref{eq:MB-2mode-unitary} is 
\begin{eqnarray}
&&\hat U_{jk} (\theta_1, \theta_2) \nonumber\\
&=& 
\hat B_{jk}^{\dag}(\pi/4) R_j \bigg(2\theta_1 + \frac{\pi}{2} \bigg)  R_k \bigg(2\theta_2 + \frac{\pi}{2} \bigg) \hat B_{jk}(\pi/4) \nonumber\\
&=&
R_j (\theta^+_{1,2}) R_k(\theta^+_{1,2}) \bigg[ R_j \bigg(\frac{\pi}{2} \bigg)  \hat B_{jk}(\theta^-_{1,2}) R_k \bigg(\frac{\pi}{2} \bigg) \bigg].
\end{eqnarray}
Therefore a variable beam splitter is implemented with some additional phase shifts.

\subsection{Measurement-based non-Gaussian unitary}\label{sec:CubicPhase}

\begin{figure*}[ht!]
\includegraphics[width=12cm]{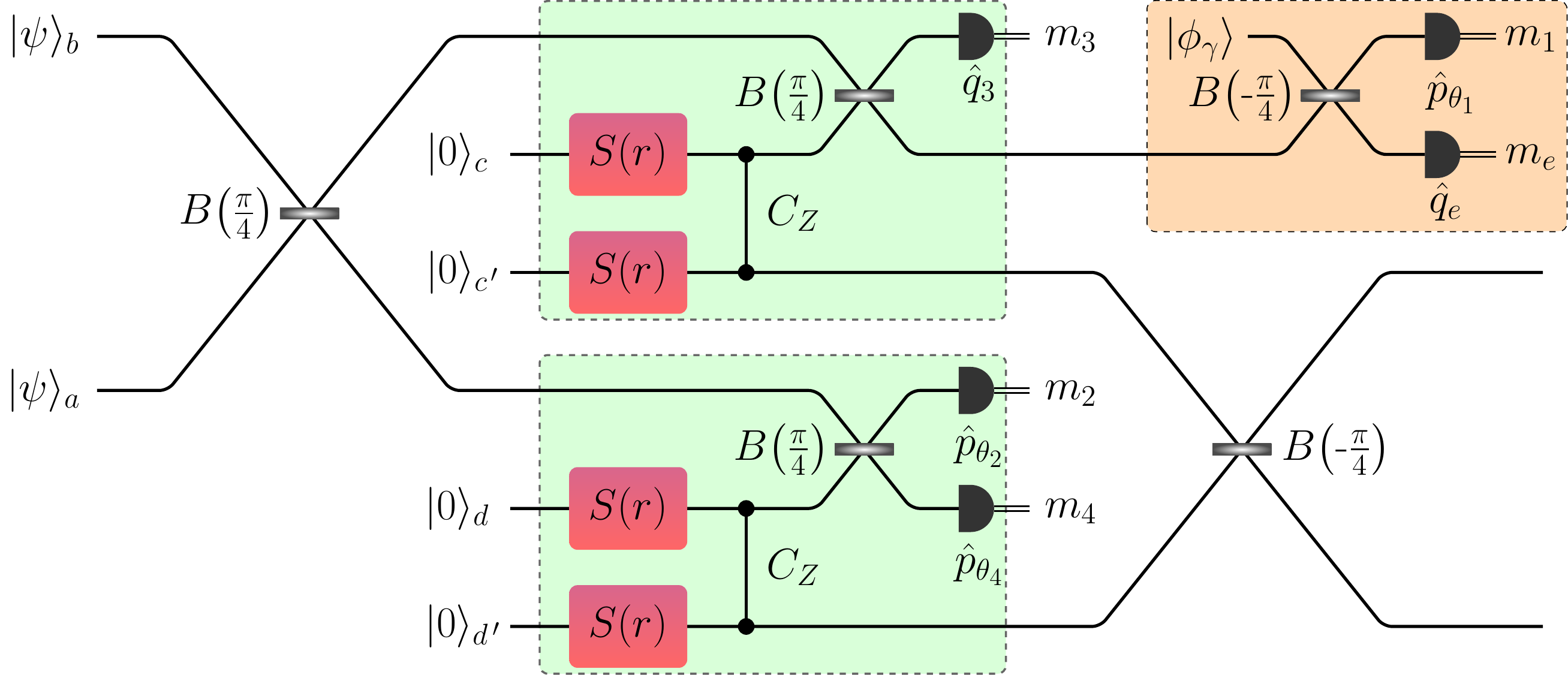}
\caption{ Implementing cubic phase gates in two-dimensional temporal cluster states. The cubic phase circuit (the orange shaded box) is fitted into a two-dimensional 
temporal cluster state in one measurement step. The whole circuit implements a cubic phase gate, a Gaussian unitary and two $50:50$ beam splitters. 
} 
\label{fig:cubic-phase-two}
\end{figure*}

For the two-dimensional temporal cluster states that we are interested in, four homodyne measurements in each step implement a two-mode Gaussian unitary 
(or two single-mode Gaussian unitaries). To include the cubic phase gate such that universal quantum computation is possible, a switch and 
a cubic phase circuit (orange shaded part in Fig. \ref{fig:cubic-phase-single}) can be introduced to the optical setup in Fig. \ref{fig:2D-cluster-setup}. When we
implement Gaussian unitaries, the switch is set to connect the homodyne detector $D_1$. When we want to implement a cubic phase gate, the switch is set to connect 
the cubic phase circuit. Fig. \ref{fig:cubic-phase-two} shows an equivalent circuit in one measurement step when implementing a cubic phase gate (and another single-mode
Gaussian unitary). 

In realistic cases, the amount of squeezing in the pulses that are used to produce the temporal cluster states is finite. This finite squeezing gives rise to errors in the implemented
unitaries. For Gaussian unitaries (and Gaussian input states), the errors due to the effect of finite squeezing can be corrected by using the information of the input states \cite{Su2018EC}. An error correction scheme for non-Gaussian gates (and states) has not been developed. Furthermore, the ideal cubic phase state is unphysical since it requires infinite energy. 
The ideal cubic phase state can be approximated by some non-Gaussian states, e.g., ON states from Sec. \ref{Sec:WeakCubicPhase} (a superposition of the vacuum and Fock states) \cite{Krishna2018ON}. In this case additional
errors are introduced. 

In next three sections we consider the implementation of three important quantum algorithms using temporal cluster states.

\section{Gaussian Boson Sampling}\label{sec:GBS}

We are now ready to discuss the implementations of quantum algorithms using two-dimensional temporal cluster states. 
We will focus on three important quantum algorithms: Gaussian Boson Sampling \cite{Hamilton2017}, CV-IQP \cite{Douce2017CVIQP} 
and CV Grover's search algorithm \cite{Pati2000}. 
To clearly illustrate the main steps of these quantum algorithms, we consider algorithms with only a few quantum modes. The generalisation to
algorithms with a large number of modes is straightforward. 

In the original implementation of the Boson Sampling algorithm \cite{Aaronson2011BS},
single photons are injected into a linear optical network, which consists of beam splitters and phase shifters, and the output state is detected by
single photon detectors to obtain the photon number distribution. The key observation is that the sampling probability of a certain output photon pattern is related to 
the permanent of the submatrices of a unitary matrix determined by the linear optical network \cite{Aaronson2011BS}. 
Given that estimating the permanent of a matrix is hard for a classical computer, while
sampling the photon number distribution can be done efficiently, it is believed that Boson Sampling may achieve quantum supremacy before a universal quantum computer is 
built \cite{Boixo2018}. 
The original Boson Sampling scheme has also been experimentally implemented \cite{Tillmann2013, Spring2013, Crespi2013, Broome2013}.

Several variants of the original Boson Sampling have been proposed. The single photon inputs can be generated by heralding the same number of two-mode squeezed vacuum states,
which is known as Scattershot Boson Sampling \cite{Lund2014SBS}. Another variant is to replace the single photon input stats by Gaussian input states \cite{Sahleh2015}. 
One important example
is to inject squeezed coherent states into a linear network and the output photon number distribution is related to the 
vibronic spectra of molecules \cite{Huh2015, {huh2017vibronic}}. Another
example is to directly inject squeezed vacuum states into the linear network \cite{Hamilton2017}. The probability of the output photon number distribution is shown to be related to
the hafnian of the submatrices of a matrix determined by the linear optical network \cite{Hamilton2017}. In particular, the number of perfect matchings of an undirected graph
can be estimated using the Gaussian Boson Sampling \cite{Bradler2017}. 

\begin{figure}[ht!]
\includegraphics[width=8.5cm]{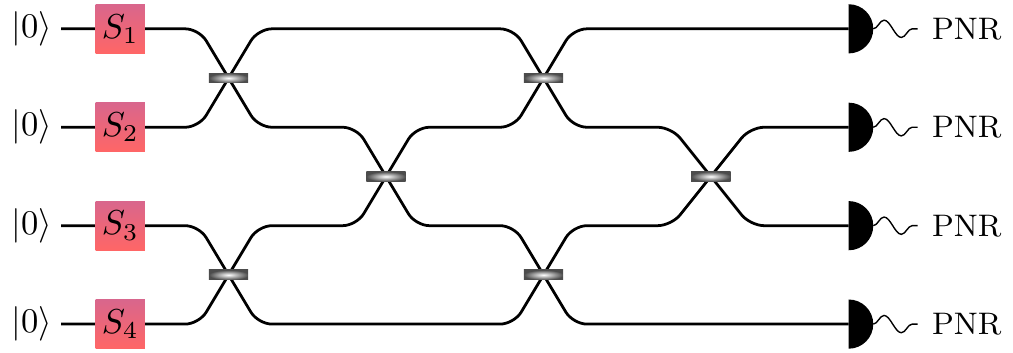}
\caption{ Gaussian Boson Sampling with four modes. We assume that all phases are included in the squeezers and beam splitters. } 
\label{fig:4mode-GBS-circuit}
\end{figure}

In contrast to the original Boson Sampling, the only non-Gaussian element in  Gaussian Boson Sampling is the photon number detection. This is due to the fact that both 
the input states and unitaries are Gaussian. In this paper, we consider the implementation of Gaussian Boson Sampling using the two-dimensional temporal 
cluster states. The implementation with linear optics using two-dimensional temporal cluster states has been discussed in Ref. \cite{Alexander2017MBLO}, as well as the original 
Boson Sampling. The main challenge there is the noise due to the effect of finite squeezing in the cluster states. We find here that the Gaussianity of the states and unitaries
in the Gaussian Boson Sampling can lead to experimental advantages because one can correct the error due to the effect of finite squeezing \cite{Su2018EC}. 

\begin{figure*}[ht!]
\includegraphics[width=10cm]{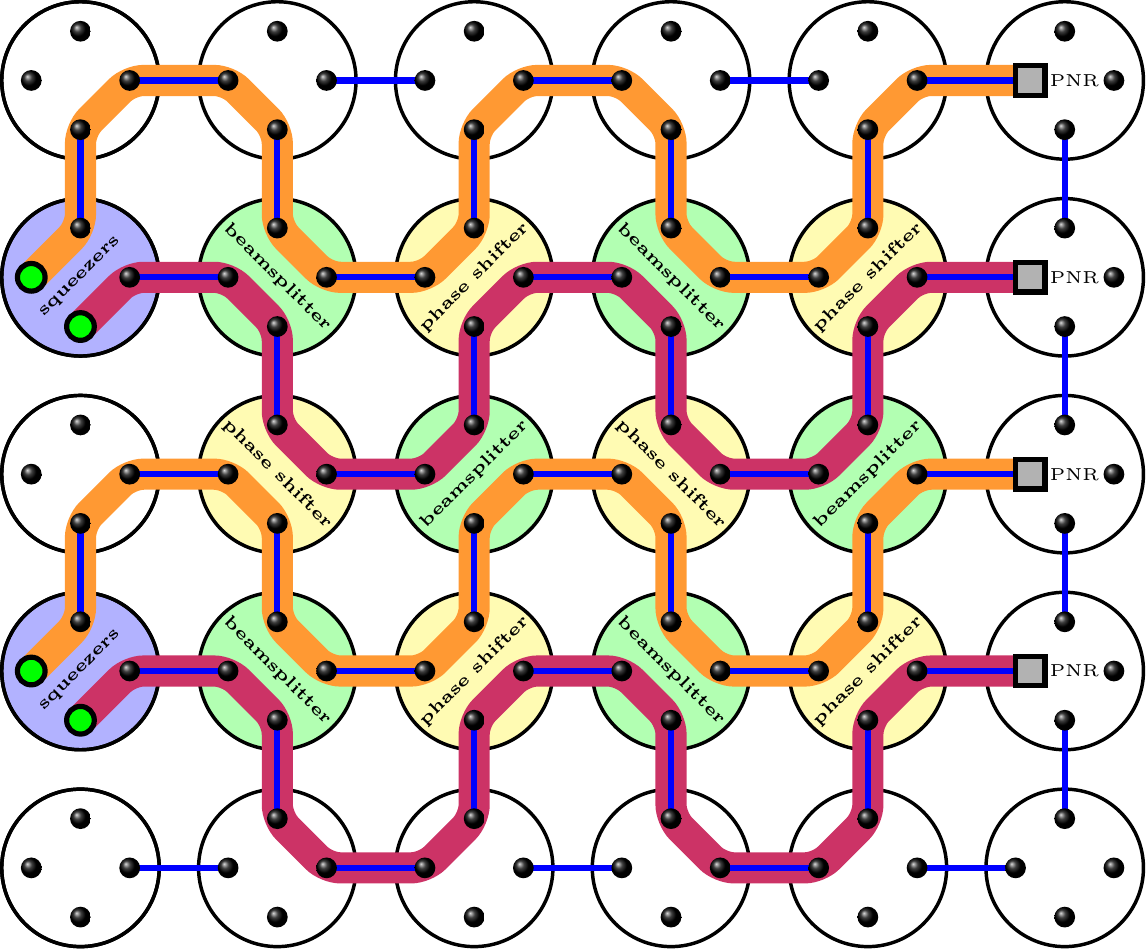}
\caption{ Measurement-based Gaussian Boson Sampling. The blue shaded macronode implements two identical single-mode squeezers, the green shaded macronode
implements a beam splitter and the yellow shaded macronode implements two phase shifts. The macronode that is not shaded implements a single-mode phase shift. } 
\label{fig:4mode-GBS-graph}
\end{figure*}

To clearly illustrate how to implement Gaussian Boson Sampling in the two-dimensional temporal cluster states, we first ignore the noise due to the effect of finite squeezing,
namely, we work in the infinite squeezing limit. Any linear optical network can be decomposed into a series of beam splitters and phase shifters \cite{clements2016optimal}, 
the implementations of which via 
measuring the cluster states have been discussed in Sec. \ref{sec:MBGU}. Given a linear optical network in the gate model, it is straightforward to map it to the
measurement-based model. Without loss of generality, we consider a Gaussian Boson Sampling with four modes. The gate model circuit is given by Fig. \ref{fig:4mode-GBS-circuit}
and the implementation in the temporal cluster states is shown in Fig. \ref{fig:4mode-GBS-graph}. In Fig. \ref{fig:4mode-GBS-graph} we assume that the input states are 
vacuum states and the squeezed states (with the same amount of squeezing) are generated via homodyne measurements. One can also prepare the input states beforehand 
and tune the amount of squeezing according to specific algorithms. 
After the linear optical network the output modes are detected by photon number resolution (PNR) detectors, giving the photon number distribution.  
Table \ref{tab:GBS} provides a detailed description of steps for the implementation of Gaussian Boson Sampling via homodyne measurements. 

Generalisation of the above implementation to a linear optical network with a large number of modes is straightforward if the squeezing is infinite. However, the 
amount of squeezing of a physical squeezed state is always finite and this results in errors when implementing a unitary via homodyne measurements \cite{Alexander2014}. 
The overall noise increases as the number of measurement steps grows. Ref. \cite{Alexander2017MBLO} showed that to implement a 
Boson Sampling with 6 photons requires about 20 dB of 
squeezing in the cluster states, which is very challenging for the state-of-the-art technologies. For measurement-based Gaussian Boson Sampling, both the states and unitaries
are Gaussian and the errors due to the effect of finite squeezing can be corrected if a moderate amount of online squeezing can be achieved \cite{Su2018EC}. Online squeezing
has been demonstrated in several experiments \cite{Miwa2014, Marshall2016} and so it is promising to include the finite-squeezing error correction on a linear network with a higher number of modes. 

\begin{table*}[tp]
\caption{Process of implementing four-mode Gaussian Boson Sampling.} 
\label{tab:GBS}\centering%
\begin{center}
    \begin{tabular}{| c | c | l |}
%
    \hline
    Time (macronode) & ~States of switches~ & Operations \\ \hline 
    $T_1$ & (${\rm \bf s}_1$, $\bar {\rm \bf s}_1$) & Homodyne measurements/Phase shift \\ \hline 
    $T_2$ & (${\rm \bf s}_2$, $\bar {\rm \bf s}_2$) & Inject input states/Homodyne measurements/Squeezing unitary \\ \hline 
    $T_3$ & (${\rm \bf s}_1$, $\bar {\rm \bf s}_1$) & Homodyne measurements/Phase shift \\ \hline 
    $T_4$ & (${\rm \bf s}_2$, $\bar {\rm \bf s}_2$) & Inject input states/Homodyne measurements/Squeezing unitary \\ \hline
    $T_5$ & (${\rm \bf s}_1$, $\bar {\rm \bf s}_1$) & None \\ \hline
    $T_6$ & (${\rm \bf s}_1$, $\bar {\rm \bf s}_1$) & Homodyne measurements/Phase shift \\ \hline 
    $T_7$ & (${\rm \bf s}_1$, $\bar {\rm \bf s}_1$) & Homodyne measurements/Beam splitter unitary \\ \hline 
    $T_8$ & (${\rm \bf s}_1$, $\bar {\rm \bf s}_1$) & Homodyne measurements/Phase shifts \\ \hline 
    $T_9$ & (${\rm \bf s}_1$, $\bar {\rm \bf s}_1$) & Homodyne measurements/Beam splitter unitary \\ \hline
    $T_{10}$ & (${\rm \bf s}_1$, $\bar {\rm \bf s}_1$) & Homodyne measurements/Phase shift \\ \hline
    $T_{11} \rightarrow T_{25}$ & (${\rm \bf s}_1$, $\bar {\rm \bf s}_1$) & Homodyne measurements/Two-mode or single-mode unitaries \\ \hline 
    $T_{26}$ & (${\rm \bf s}_1$, $\bar {\rm \bf s}_3$) & Readout output state/Photon number detection \\ \hline 
    $T_{27}$ & (${\rm \bf s}_1$, $\bar {\rm \bf s}_3$) & Readout output state/Photon number detection \\ \hline 
    $T_{28}$ & (${\rm \bf s}_1$, $\bar {\rm \bf s}_3$) & Readout output state/Photon number detection \\ \hline
    $T_{29}$ & (${\rm \bf s}_1$, $\bar {\rm \bf s}_3$) & Readout output state/Photon number detection \\
    \hline
    \end{tabular}
\end{center}
\end{table*}

\section{Continuous-variable Instantaneous Quantum Polynomial circuits}\label{sec:IQP}

The instantaneous quantum polynomial (IQP) computation is a particular kind of quantum computation that consists of only commuting gates. It was shown 
that sampling the output probability distributions of the IQP computation cannot be efficiently achieved by classical computers \cite{Bremner2011}. IQP circuits
have been extended to the continuous-variable domain, denoted as CV-IQP, by using squeezed states and homodyne detection \cite{Douce2017CVIQP}. 
A particular implementation of a CV-IQP circuit is to include unitaries that are only functions of position quadratures, e.g., $e^{i f(\hat x)}$ where the function
$f(\hat x)$ is an arbitrary polynomial of $\hat x$. If $f(\hat x)$ is a polynomial up to a quadratic function of $\hat x$, then the unitary is Gaussian and can be implemented
efficiently. If $f(\hat x)$ is a cubic function of $\hat x$, then the unitary represents a cubic phase gate, the implementation of which has been discussed in detail in 
Sec. \ref{sec:CubicPhase}. If $f(\hat x)$ is a higher-order polynomial of $\hat x$, the direct implementation of the unitary is in principle possible but is 
challenging \cite{Krishna2018ON}. The strategy is to decompose the unitary into Gaussian gates and cubic phase gates \cite{BraunsteinLloyd1999}. The decomposition 
requires unitaries that are also functions of $\hat p$, e.g., the Fourier transform ${F}$. Therefore, after the decomposition the equivalent circuit does not look like 
a CV-IQP circuit due to the presence of non-commuting unitaries. However, if each higher-order unitary is considered as a whole as if it wasn't decomposed, 
then the circuit is still a CV-IQP circuit.

\begin{figure*}[!tbp]
\centering
\subfloat[]{%
  \includegraphics[clip,width=8.3cm]{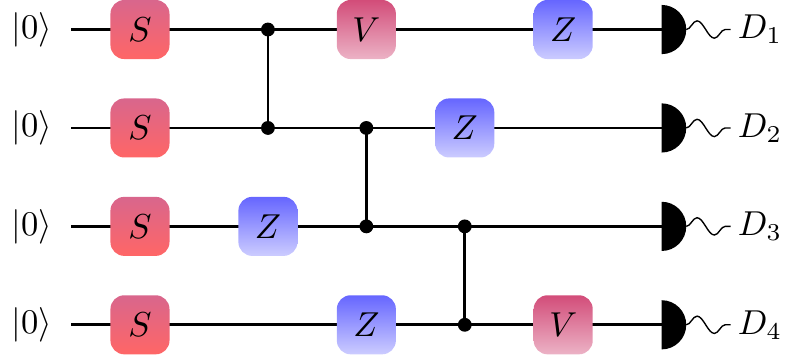}%
}
\hfill
\subfloat[]{%
  \includegraphics[clip,width=8.3cm]{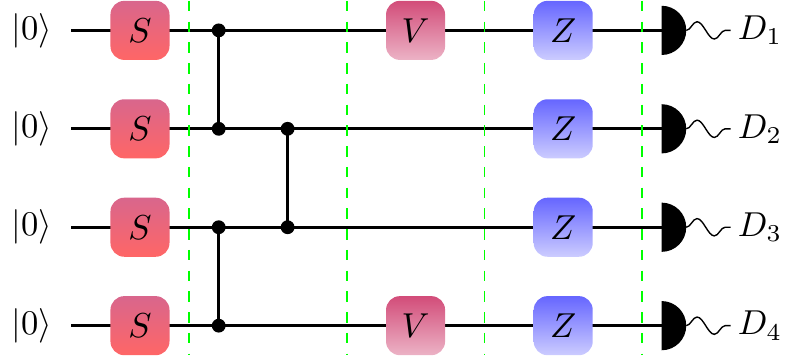}%
}

\caption{ CV-IQP circuit. (a) An example of a four-mode CV-IQP circuit. The four squeezers generate four single-mode momentum-squeezed states. The commuting gates
include displacement $Z$, cubic phase gate $V$ and controlled-Z gate $C_Z$. The output modes are detected by homodyne detectors $D_i$, $i=1,..,4$. 
(b) The commuting gates are rearranged such that all controlled-Z gates are right after the single-mode squeezers, all displacements are right before the homodyne detectors
and the cubic phase gates are in between. }
\label{fig:CV-IQP}
\end{figure*}

The  CV-IQP circuit can be decomposed into three parts \cite{Arrazola2017}: momentum-squeezed vacuum states as inputs, a sequence of commuting unitaries
and homodyne detectors. Fig. \ref{fig:CV-IQP} (a) shows an example of a typical four-mode CV-IQP circuit up to third order unitaries. 
In this paper, we are interested in implementing the CV-IQP circuits using the two-dimensional temporal cluster states and homodyne measurements. It is straightforward to 
conceive a measurement-based implementation given a CV-IQP circuit like Fig. \ref{fig:CV-IQP} (a). However, we find it very helpful to rearrange those commuting unitaries. 
The strategy is to move all controlled-Z gates $C_Z$ to be directly after the squeezed vacuum states and move all momentum displacement operators $Z$ to be directly before the homodyne 
detectors, as shown in Fig. \ref{fig:CV-IQP} (b). The momentum displacements can be absorbed into the feedforward which we have to do before the homodyne measurements, 
therefore we only need to consider a series of controlled-Z gates, which are Gaussian unitaries, and a series of cubic-phase gates. The main advantage of this rearrangement is 
that before the action of the cubic phase gates, both the state and unitaries are Gaussian. When implementing the controlled-Z gates using cluster states, we can correct the errors
due to the effect of finite squeezing \cite{Su2018EC}. Since the finite squeezing noise increases as the number of implemented gates increases, this rearrangement therefore can
significantly reduce the noise. 

\begin{figure*}[ht!]
\includegraphics[width=14cm]{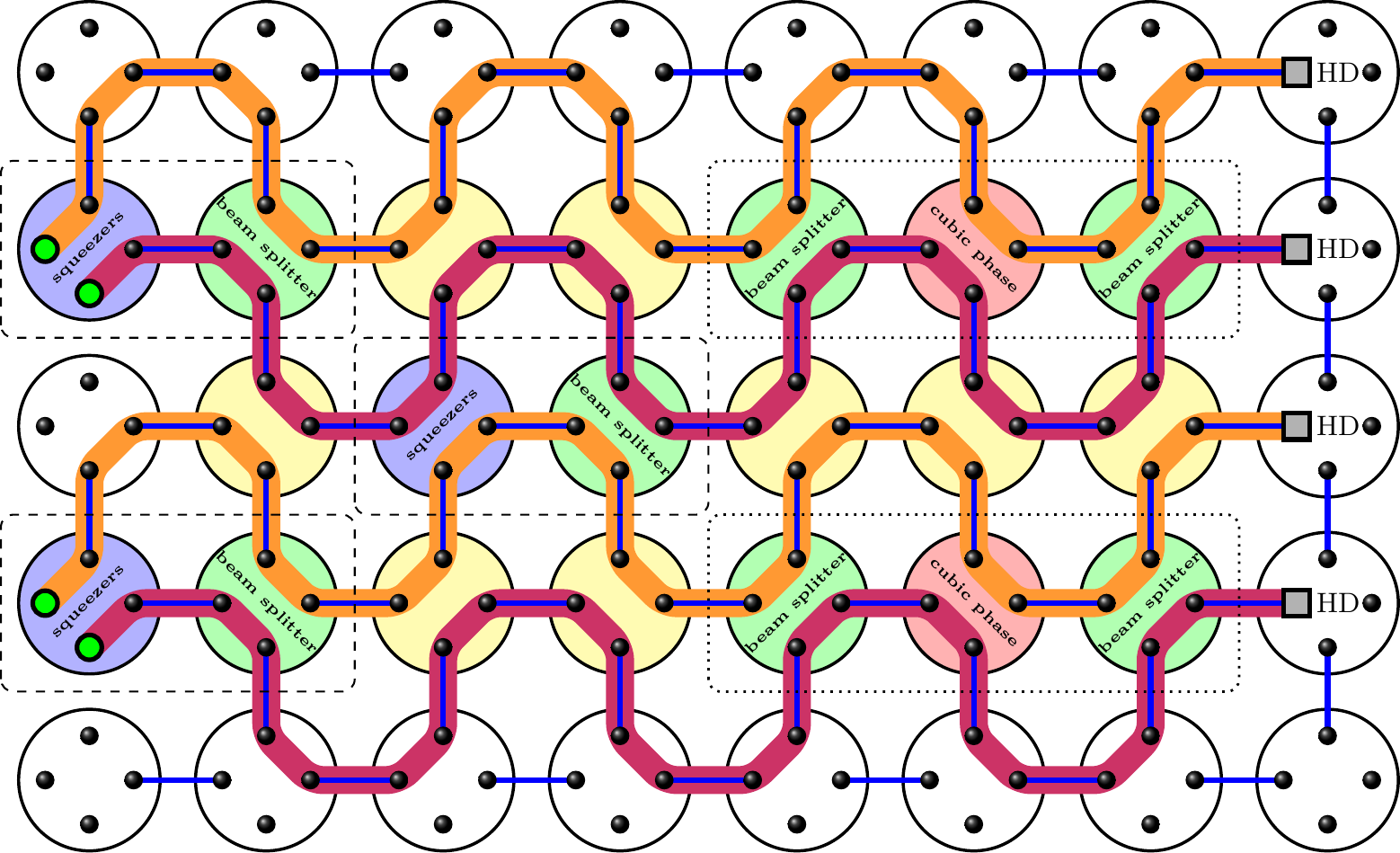}
\caption{ Measurement-based four-mode CV-IQP circuit. The blue shaded macronode implements two identical single-mode squeezers, the green shaded macronode
implements a beam splitter (basically a $50:50$ beam splitter),  the pink shaded macronode implements a cubic phase gate and an identity gate, the yellow shaded 
macronode implements two phase shifts which is not explicitly labelled. The macronode that implements a single mode gate is not filled with color, which we assume 
implements a phase shift. In the first column of macronodes, the four input modes (green circles) are prepared in the momentum squeezed states. In the first four columns,
the two macronodes in a black dashed box together implements a controlled-Z gate. From the fifth to the seventh column, the three macronodes in a black dotted box together
implements a cubic phase gate and an identity gate. In the last column, the four output modes (grey squares) are detected by homodyne detectors. } 
\label{fig:4mode-IQP-graph}
\end{figure*}

A direct implementation of the controlled-Z gate is challenging, however it can be decomposed into two single-mode squeezers with equal squeezing parameters and squeezing angles,
and a beam splitter \cite{Loock2007}. Two identical single-mode squeezers and a beam splitter can be implemented directly via homodyne measurements,
as discussed in Sec. \ref{sec:MBGU}. Fig. \ref{fig:4mode-IQP-graph} shows an implementation of the four-mode CV-IQP circuit shown in Fig. \ref{fig:CV-IQP} via measuring
the two-dimensional temporal cluster states. 
Table \ref{tab:CV-IQP} describes the detailed process of implementing the CV-IQP circuit of Fig. \ref{fig:CV-IQP}. If higher-order unitaries are required, e.g., $e^{i \lambda \hat x^4}$, 
we have to first decompose it into series of Gaussian gates and cubic phase gates and then implement it just as in Fig. \ref{fig:4mode-IQP-graph}. Generalisation to a CV-IQP
circuit with a larger number of modes is straightforward. 

\begin{table*}[tp]
\caption{Process of implementing four-mode CV-IQP circuit.} 
\label{tab:CV-IQP}\centering%
\begin{center}
    \begin{tabular}{| c | c | l |}
    \hline
    Time (macronode) & ~States of switches~ & Operations \\ \hline 
    $T_1$ & ($\boldsymbol{\rm s}_1$, $\bar {\boldsymbol{\rm s}}_1$) & Homodyne measurements/Phase shift \\ \hline 
    $T_2$ & ($\boldsymbol{\rm s}_2$, $\bar {\boldsymbol{\rm s}}_2$) & Inject input states/Homodyne measurements/Squeezing unitary \\ \hline 
    $T_3$ & ($\boldsymbol{\rm s}_1$, $\bar {\boldsymbol {\rm s}}_1$) & Homodyne measurements/Phase shift \\ \hline 
    $T_4$ & ($\boldsymbol{\rm s}_2$, $\bar {\boldsymbol{\rm s}}_2$) & Inject input states/Homodyne measurements/Squeezing unitary \\ \hline
    $T_5$ & ($\boldsymbol{\rm s}_1$, $\bar {\boldsymbol{\rm s}}_1$) & None \\ \hline
    $T_6 \rightarrow T_{10}$ & ($\boldsymbol{\rm s}_1$, $\bar {\boldsymbol{\rm s}}_1$) & Two beam splitters and phase shifts \\ \hline 
    $T_{11} \rightarrow T_{15}$ & ($\boldsymbol{\rm s}_1$, $\bar {\boldsymbol{\rm s}}_1$) & Squeezers and phase shifts \\ \hline 
    $T_{15} \rightarrow T_{20}$ & ($\boldsymbol{\rm s}_1$, $\bar {\boldsymbol{\rm s}}_1$) & One beam splitter and phase shifts \\ \hline 
    $T_{21} \rightarrow T_{25}$ & ($\boldsymbol{\rm s}_1$, $\bar {\boldsymbol{\rm s}}_1$) & Two beam splitters and phase shifts \\ \hline
    $T_{26}$ & ($\boldsymbol{\rm s}_1$, $\bar{\boldsymbol {\rm s}}_1$) & Homodyne measurements/Phase shift \\ \hline
    $T_{27}$ & ($\boldsymbol{\rm s}_1$, $\bar {\boldsymbol{\rm s}}_1$) & Inject cubic phase state/Homodyne measurements/Cubic phase gate \\ \hline
    $T_{28}$ & ($\boldsymbol{\rm s}_1$, $\bar {\boldsymbol{\rm s}}_1$) & Homodyne measurements/Phase shifts \\ \hline
    $T_{29}$ & ($\boldsymbol{\rm s}_1$, $\bar {\boldsymbol{\rm s}}_1$) & Inject cubic phase state/Homodyne measurements/Cubic phase gate  \\ \hline
    $T_{30}$ & ($\boldsymbol{\rm s}_1$, $\bar {\boldsymbol{\rm s}}_1$) & Homodyne measurements/Phase shift \\ \hline
    $T_{31} \rightarrow T_{35}$ & ($\boldsymbol{\rm s}_1$, $\bar {\boldsymbol{\rm s}}_1$) & Two beam splitters and phase shifts \\ \hline 
    $T_{36}$ & ($\boldsymbol{\rm s}_1$, $\bar {\boldsymbol{\rm s}}_3$) & Readout output state/Homodyne detection \\ \hline 
    $T_{37}$ & ($\boldsymbol{\rm s}_1$, $\bar {\boldsymbol{\rm s}}_3$) & Readout output state/Homodyne detection \\ \hline 
    $T_{38}$ & ($\boldsymbol{\rm s}_1$, $\bar {\boldsymbol{\rm s}}_3$) & Readout output state/Homodyne detection \\ \hline
    $T_{39}$ & ($\boldsymbol{\rm s}_1$, $\bar {\boldsymbol{\rm s}}_3$) & Readout output state/Homodyne detection \\
    \hline
    \end{tabular}
\end{center}
\end{table*}

\section{Continuous-variable Grover search algorithm}\label{sec:CVGrover}

One of the first quantum algorithms to show a speed up compared to any classical counterpart was Grover's quantum search algorithm \cite{Grover1997}. In this algorithm an unsorted list of $N$ entries could be searched to find an unmarked item with $O(\sqrt{N})$ steps, instead of the best classical case which requires $O(N)$ steps. This quantum search algorithm, originally proposed for discrete-variable quantum systems~\cite{Grover1997}, was generalised to continuous variables by Pati \textit{et al.} \cite{Pati2000}. It was argued that the CV proposal may be superior to any discrete-variable implementation for large database searches, due to the considerable amount of information that could in principle be encoded with small CV systems. 

\begin{figure}[ht!]
\includegraphics[width=\columnwidth]{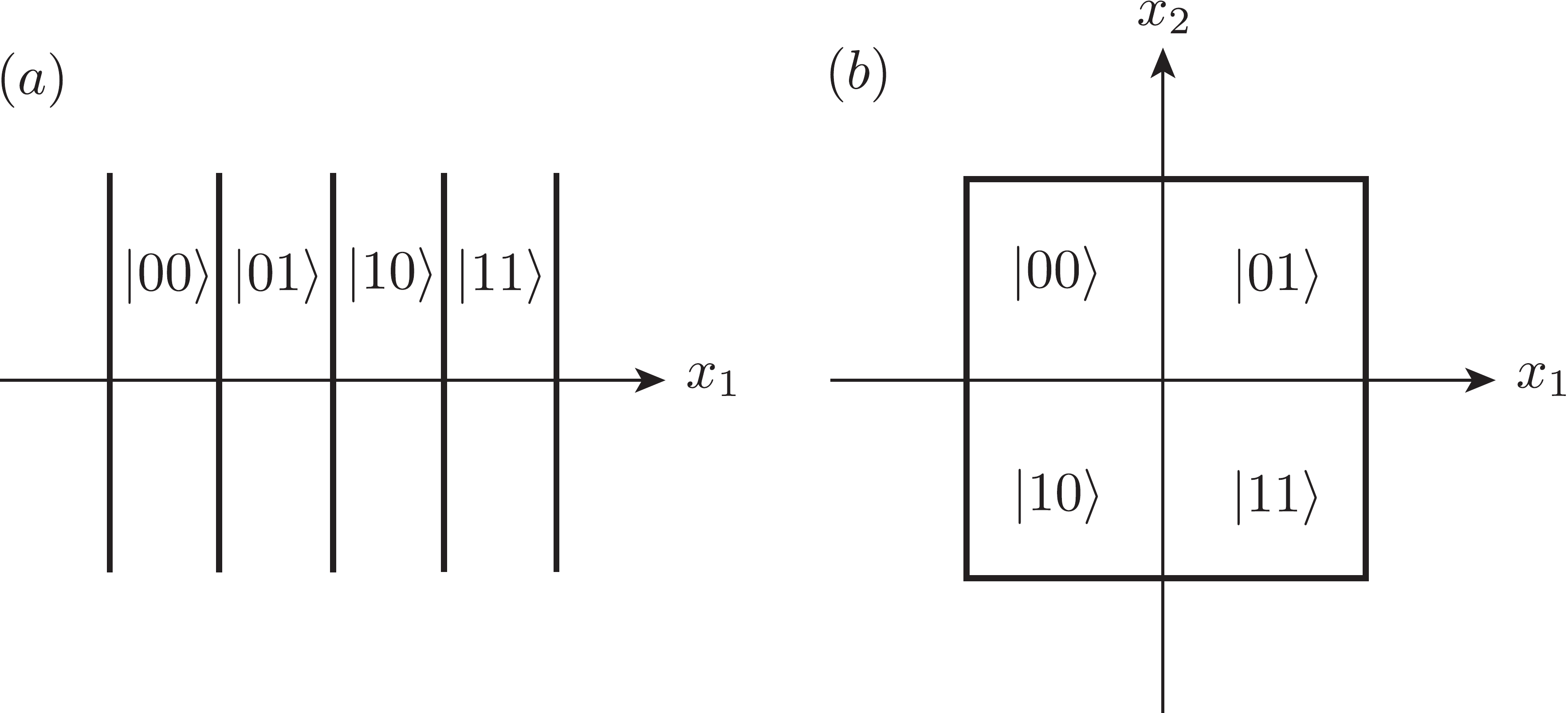}
\caption{Geometry for encoding $N$ qubits into $n$ continuous variables: (a) State space division for  $N=4$ and $n=1$; (b) State space division for  $N=4$ and $n=2$.  } 
\label{fig:Pati1D2D}
\end{figure}

In the original CV quantum search algorithm proposal \cite{Pati2000}, the information of a database with $N$ entries $\{1, 2, \ldots, N\}$ is encoded into $n$ continuous variables or modes state $\ket{x}=\ket{x_{1}, x_{2}, \ldots, x_{n}}$, by dividing a compact subspace of the $n$-dimensional state space into $N$ equal subvolumes $\Delta x$. 
In the case when $N=4$ and $n=1$, the one-dimensional state space will be divided into four equal regions, as shown in Fig. \ref{fig:Pati1D2D}~(a); if $N=4$ and $n=2$, the two-dimensional state space will be divided into four equal regions, as shown in Fig. \ref{fig:Pati1D2D}~(b). An alternative method of encoding information into a single continuous variable was presented by Adcock~\textit{et al.} \cite{Adcock2009}, in the context of the Deutsch-Jozsa algorithm. In this case each subregion in the one-dimensional state space corresponds to a single bit of information. In this paper we use the encoding of Pati \textit{et al.} \cite{Pati2000}.    

\begin{figure}[ht!]
\includegraphics[width=\columnwidth]{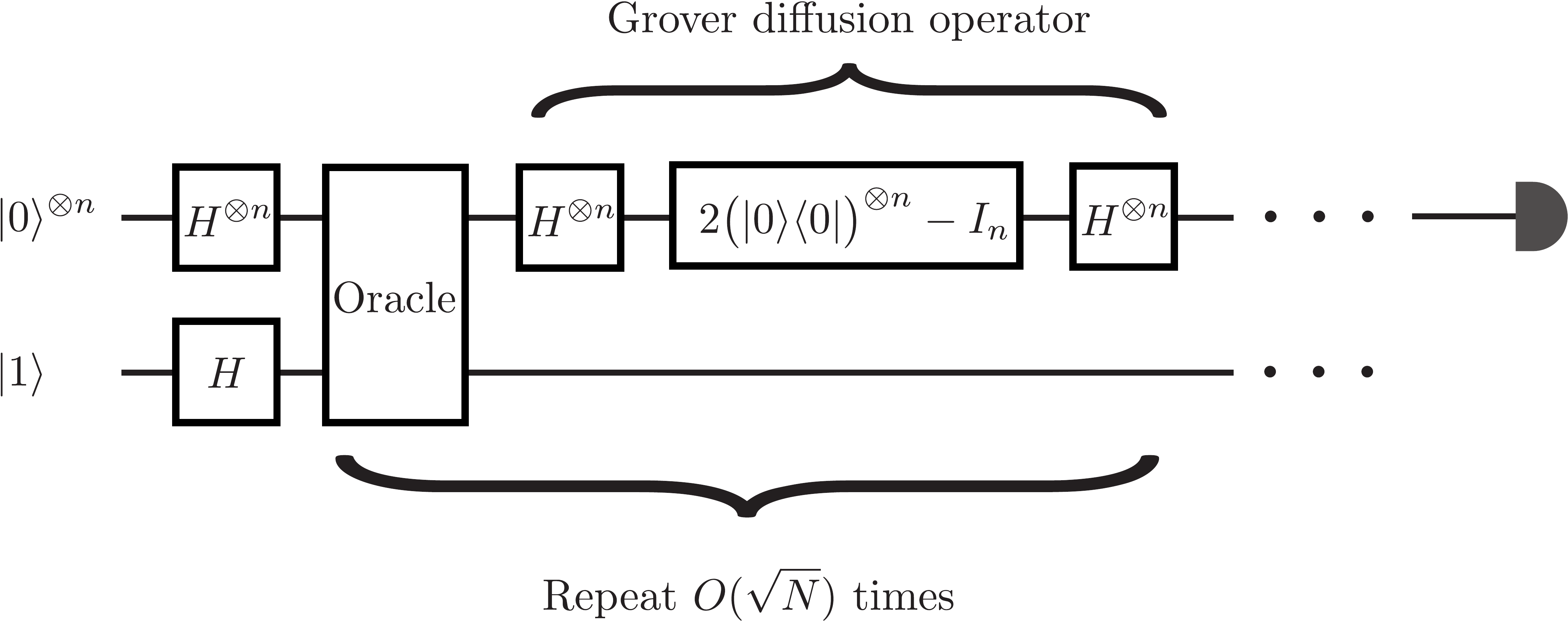}
\caption{Circuit diagram for Grover's search algorithm~\cite{Grover1997, NielsenBook}. } 
\label{fig:GroverCircuit}
\end{figure}

The basic elements required for Grover's search algorithm with discrete-variables~\cite{NielsenBook}, shown in Fig. \ref{fig:GroverCircuit}, are an oracle, which recognises solutions to the searching problem, Hadamard gates and the Grover diffusion operator. We assume that the oracle is a given black box operation and concentrate on the Hadamard gate and Grover diffusion operator. In continuous variables the analog to the Hadamard gate is the Fourier gate \cite{Pati2000}, which can easily be implemented optically with a $\pi/2$ phase shift, where ${F}\ket{x}$ is an eigenstate of the conjugate quadrature. The action of the Fourier gate is
\begin{equation}
{F}\ket{x}=\frac{1}{\sqrt{\pi^{n}}}\int \mathrm d y ~e^{2ixy} \ket{y},
\end{equation}
where $xy=x_{1}y_{1}+\cdots +x_{n}y_{n}$, $\ket{y}=\ket{y_{1},y_{2},\cdots , y_{n}}$ and both $x$ and $y$ are in the position basis. 

In continuous variables the Grover diffusion operator is a selective inversion operator \cite{Pati2000} defined as 
\begin{equation}
\hat{I}_{x}=\mathbb{I}-2P_{\Delta x}, 
\end{equation}
where $\mathbb{I}$ is the identity operator and $P_{\Delta x}$ is the projection operator defined as
\begin{equation}
P_{\Delta x}=\int_{x_{0}-\Delta x/2}^{x_{0}+\Delta x/2} \mathrm d x' \ket{x'}\bra{x'}\label{Eqn:PDeltax},
\end{equation}
defining a projection operator for a subvolume $\Delta x$ centered at $x_0$. The main challenge with implementing a CV version of Grover's search algorithm is the implementation of this selective inversion operator $\hat{I}_{x}$. 

The obvious first attempt to implement this selective inversion operator $\hat{I}_{x}$ would be to use the results of Lloyd and Braunstein \cite{BraunsteinLloyd1999}: any Hamiltonian consisting of an arbitrary polynomial of the conjugate operators $\hat{x}$ and $\hat{p}$ can be constructed with only Gaussian operations and a single operator with order greater than 2 in $\hat{x}$ or $\hat{p}$, such as the cubic phase gate. The selective inversion operator can be rewritten as $\hat{I}_{x}=e^{\ln(\mathbb{I}-2P_{\Delta x})}=e^{P_{\Delta x}\ln(-1)}=e^{i\pi P_{\Delta x}}$. However it is not immediately obvious that the projector $P_{\Delta x}$ can be written as a polynomial of $\hat{x}$ and $\hat{p}$ operators. 

If we consider a particular geometrical representation for the selective inversion operator $\hat{I}_{x}$ by choosing the number of encoding continuous variables $n$, we can write $\hat{I}_{x}$ as a function in state space:  $\hat{I}_{x} = f(\hat{x}_{1}, \ldots, \hat{x}_{n})$. The form of this function is a direct consequence of the projector $P_{\Delta x}$: it is a step function with positive constant values for $N-1$ database entries/state space regions and a negative constant value for the flagged region. 
The operator $P_{\Delta x}$ can be expressed as a function of $f(\hat{x}_{1}, \ldots, \hat{x}_{n})$: $P_{\Delta x}=\frac{1}{2}(\mathbb{I}-f(\hat{x}_{1}, \ldots, \hat{x}_{n}))$. Since $\hat{I}_{x}=e^{i\pi P_{\Delta x}}$, we can re-write the relation between $\hat{I}_{x}$ and $f(\hat{x}_{1}, \ldots, \hat{x}_{n})$ as:  
\begin{equation}
\hat{I}_{x}=i e^{-i\frac{\pi}{2}f(\hat{x}_{1}, \ldots, \hat{x}_{n})}.\label{eqn:Ifun}
\end{equation}

\begin{figure}[ht!]
\includegraphics[width=\columnwidth]{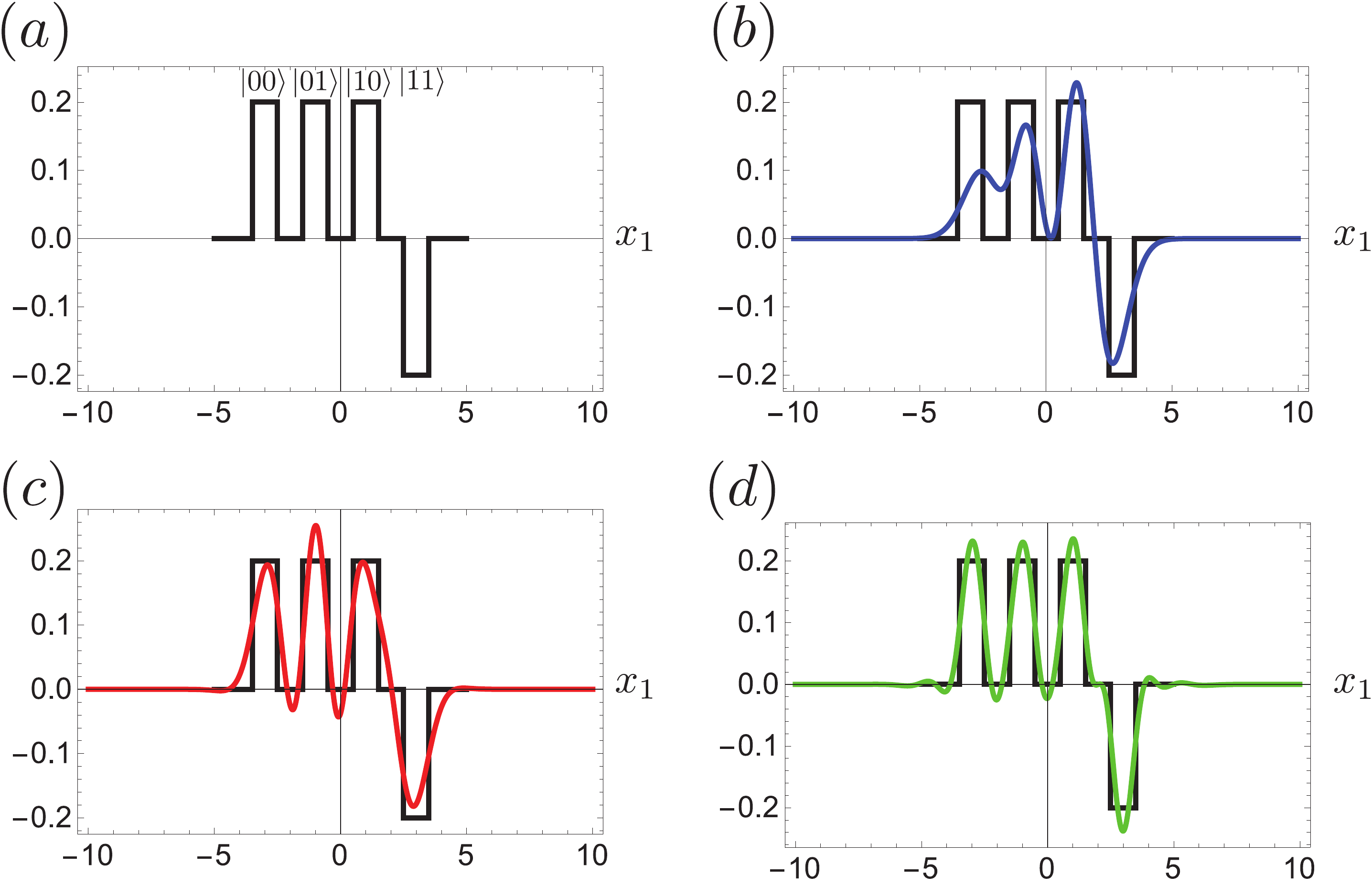}
\caption{(a) Selective inversion operator $\hat{I}_{x}$ for $N=4$ and $n=1$. $\hat{I}_{x}$ function $f(x_{1})$ expanded in the Fock basis states up to (b) 5 photons;  (c) 10 photons;   (d)  20 photons.} 
\label{fig:Pati1DFun}
\end{figure}

\begin{figure}[ht!]
\includegraphics[width=\columnwidth]{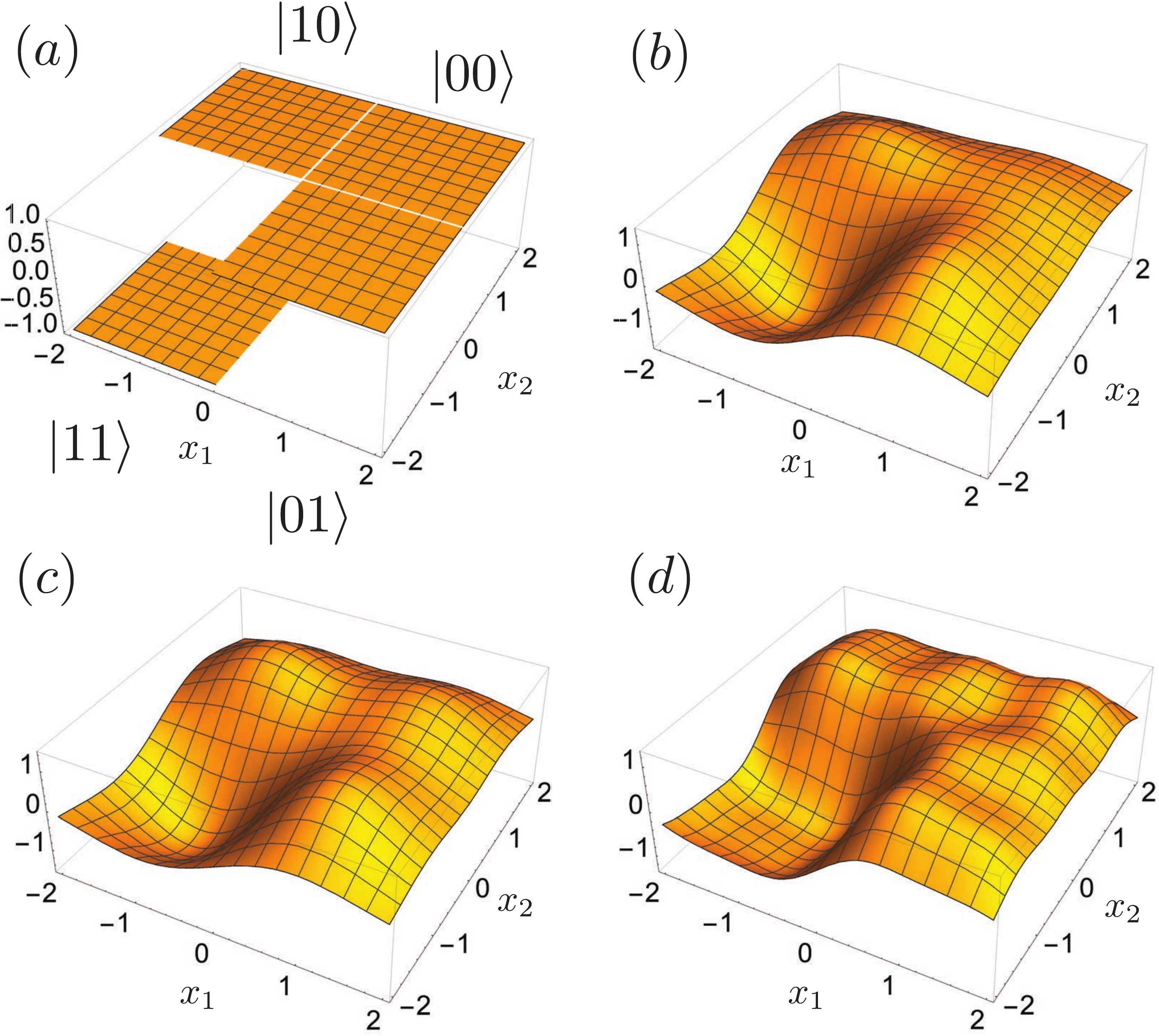}
\caption{(a) Selective inversion operator $\hat{I}_{x}$ for $N=4$ and $n=2$. $\hat{I}_{x}$ function $f(x_{1}, x_{2})$ expanded in the Fock basis states up to (b) 3 photons ; (c) 5 photons; (d) 10 photons in each mode.} 
\label{fig:Pati2DFun}
\end{figure}

The function $f(\hat{x}_{1}, \ldots, \hat{x}_{n})$ is not necessarily a polynomial function in state space variables $\hat{x}_{1}, \ldots, \hat{x}_{n}$. In the case that $n=1$, $\hat{I}_{x}$ will be a top hat function, shown in Fig. \ref{fig:Pati1DFun}~(a). In the case that $n=2$, $\hat{I}_{x}$ will be a two-dimensional step function, shown in Fig. \ref{fig:Pati2DFun}~(a).  If the function $f(\hat{x}_{1}, \ldots, \hat{x}_{n})$ could be decomposed into a polynomial function of position operators $\hat{x}_{i}$, then we could implement the operator $\hat{I}_{x}$ via the gate teleportation circuits, similar to that in Fig. \ref{fig:cubic-phase-single}. The gate teleportation would then reduce to a state preparation problem. 

We assume the function $f(\hat{x}_{1}, \ldots, \hat{x}_{n})$ is a well behaved function such that $f(\hat{x}_{1}, \ldots, \hat{x}_{n})\ket{x_{1},\ldots, x_{n}}=f(x_{1}, \ldots, x_{n})\ket{x_{1},\ldots, x_{n}}$ and expand it as a sum over the orthonormal Fock state wavefuctions $\psi_{n}(x_{i})=\langle  x_{i}\ket{n}$~\cite{ArfkenBook}: 
\begin{equation}
f(x_{1}, \ldots, x_{n})=\sum_{k_{1}, \ldots, k_{n}=0}^{\infty}c_{k_{1}, \ldots, k_{n}}\psi_{k_{1}}(x_{1})\cdots \psi_{k_{n}}(x_{n})\label{Eqn:function}.
\end{equation}
Given the chosen geometry resulting from the number of encoding continuous variables $n$, we can truncate the sum in Eq. \eqref{Eqn:function} to a ensure convergence for the given number of entires to be searched over $N$ . If we consider the $n=1$ case, the function becomes
\begin{align}
f(x) &=\sum_{k=0}^{\infty}c_{k}\psi_{n}(x) \nonumber\\
&\approx \sum_{k=0}^{m}\left(\sqrt{\pi}2^{k}k!\right)^{-\frac{1}{2}}e^{-\frac{x^{2}}{2}}\mathcal{H}_{k}(x)\nonumber\\
&\approx \sum_{j=0}^{p}\sum_{k=0}^{m}\frac{1}{\sqrt{\sqrt{\pi}2^{k}k!}j!}\biggl(\frac{-x^{2}}{2}\biggr)^{j}\mathcal{H}_{k}(x)\label{eqn:functionpoly},
\end{align}
where $m$ is the order of the highest Fock state to retain and $\mathcal{H}_{k}(x)$ is the Hermite polynomial, a known polynomial function in $x$ for a given $k$. In the last equality, we have expanded the Gaussian function $e^{-x^2/2}$ and truncated at the order $x^{2p}$, such that a significantly good approximation is obtained.
Given the truncation of the sum in Eq. \eqref{eqn:functionpoly} to $m p$ terms, there will only be terms of order $x^{m+2p}$ and lower. When we combine this with Eq. \eqref{eqn:Ifun} this implies that the selective inversion operator $\hat{I}_{x}$ will take the form $\hat{I}_{x}\approx i e^{-i\frac{\pi}{2}\sum_{q=0}^{m+2p}c_{q}\hat{x}^{q}}$. From this point it is clear that we can decompose $\hat{I}_{x}$ into a sequence of higher-order quadrature phase gates \cite{h-order}. 

\begin{figure}[ht!]
\includegraphics[width=\columnwidth]{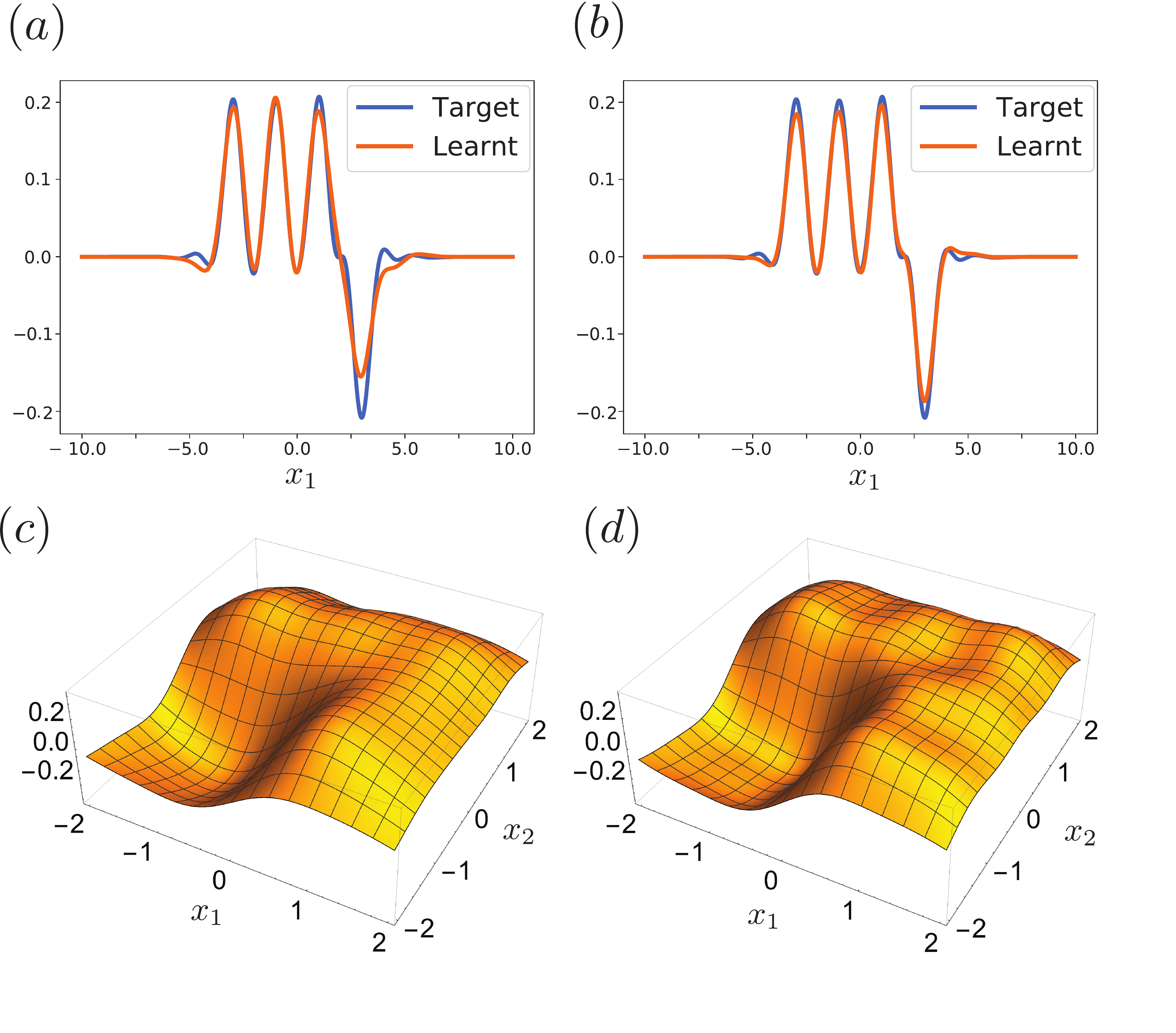}
\caption{In (a) $\&$ (b) we use the photonic quantum information optimisation routine StrawberryFields~\cite{SFpaper} to prepare the 20 photon state  $\ket{\psi_{\text{prep}}}_{1}$ (Fig.~\ref{fig:Pati1DFun}~(d)). In (c) $\&$ (d) we show the results of preparing the 10 photon per mode state $\ket{\psi_{\text{prep}}}_{2}$ (Fig. \ref{fig:Pati2DFun}~(d)). (a) Layer depth of 6 and 1000 optimisation repetitions, final fidelity = $95.7\%$; (b) Layer depth of 12 and 10000 optimisation repetitions, final fidelity = $99.4\%$; (c) Layer depth of 6 and 5000 optimisation repetitions, final fidelity = $96.3\%$; (d) Layer depth of 12 and 10000 optimisation repetitions, final fidelity = $99.7\%$. One layer consists of the gate sequence: Displacement-Rotation-Squeezer-Kerr. } 
\label{fig:1D2DSF}
\end{figure}

\begin{figure*}[ht!]
\includegraphics[width=14cm]{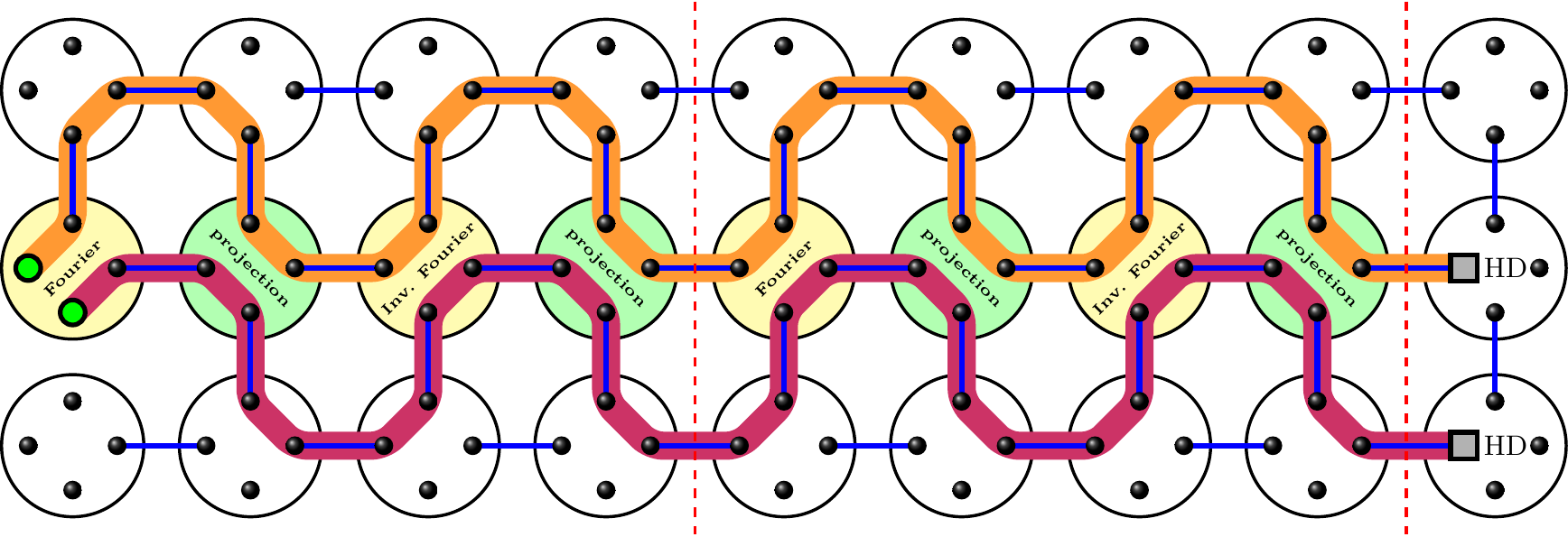}
\caption{Measurement-based two mode CV-Grover search algorithm. The yellow shaded macronodes implement Fourier or inverse Fourier transforms and the green shaded macronodes implement the selective inversion operator $\hat{I}_{x}$. The input states (two green circles) are two independent single-mode squeezed states 
(squeezed in $x$ quadrature), which are the approximations of the position eigenstates. After two applications of $\mathcal{C}$ (separated by two red dashed lines), 
the outputs are detected by homodyne detectors. 
In this case $N=4$, so four applications of the selective inversion operator are required~\cite{Pati2000}. } 
\label{fig:GroverTemporal}
\end{figure*}

\begin{table*}[tp]
\caption{Process of implementing two-mode Grover's search algorithm.} 
\label{tab:CV-Grover}\centering%
\begin{center}
    \begin{tabular}{| c | c | l |}
    \hline
    Time (macronode) & ~States of switches~ & Operations \\ \hline 
    $T_1$ & ($\boldsymbol{\rm s}_1$, $\bar {\boldsymbol{\rm s}}_1$) & Homodyne measurements/Phase shift \\ \hline 
    $T_2$ & ($\boldsymbol{\rm s}_2$, $\bar {\boldsymbol{\rm s}}_2$) & Inject input states/Homodyne measurements/Fourier transform \\ \hline 
    $T_3$ & ($\boldsymbol{\rm s}_1$, $\bar {\boldsymbol {\rm s}}_1$) & None \\ \hline 
    $T_4 \rightarrow T_{6}$ & ($\boldsymbol{\rm s}_1$, $\bar {\boldsymbol{\rm s}}_1$) & Projection and phase shifts \\ \hline 
    $T_{7} \rightarrow T_{9}$ & ($\boldsymbol{\rm s}_1$, $\bar {\boldsymbol{\rm s}}_1$) & Inverse Fourier transfom and phase shifts \\ \hline 
    $T_{10} \rightarrow T_{12}$ & ($\boldsymbol{\rm s}_1$, $\bar {\boldsymbol{\rm s}}_1$) & Projection and phase shifts \\ \hline 
    $T_{13} \rightarrow T_{15}$ & ($\boldsymbol{\rm s}_1$, $\bar {\boldsymbol{\rm s}}_1$) & Fourier transform and phase shifts \\ \hline
    $T_{16} \rightarrow T_{18}$ & ($\boldsymbol{\rm s}_1$, $\bar {\boldsymbol{\rm s}}_1$) & Projection and phase shifts \\ \hline
    $T_{19} \rightarrow T_{21}$ & ($\boldsymbol{\rm s}_1$, $\bar {\boldsymbol{\rm s}}_1$) & Inverse Fourier transform and phase shifts \\ \hline
    $T_{22} \rightarrow T_{24}$ & ($\boldsymbol{\rm s}_1$, $\bar {\boldsymbol{\rm s}}_1$) & Projection and phase shifts \\ \hline
    $T_{25}$ & ($\boldsymbol{\rm s}_1$, $\bar {\boldsymbol{\rm s}}_1$) & None \\ \hline 
    $T_{26}$ & ($\boldsymbol{\rm s}_1$, $\bar {\boldsymbol{\rm s}}_3$) & Readout output state/Homodyne detection \\ \hline
    $T_{27}$ & ($\boldsymbol{\rm s}_1$, $\bar {\boldsymbol{\rm s}}_3$) & Readout output state/Homodyne detection \\
    \hline
    \end{tabular}
\end{center}
\end{table*}

We show two explicit examples for approximating the selective inversion operator $\hat{I}_{x}$: the one dimensional case $f(x_{1})$ in Fig. \ref{fig:Pati1DFun} and the two dimensional case $f(x_{1}, x_{2})$ in Fig. \ref{fig:Pati2DFun}. The state required to be produced for the gate teleportation can be written as a sum of Fock states $\ket{\psi_{\text{prep}}}_{1}=\sum_{k=0}^{m}c_{k}\ket{k}$ for the one-dimensional case. In Fig. \ref{fig:Pati1DFun} we can see how well the approximation in Eq.~(\ref{eqn:functionpoly}) holds for the $m=5$, 10 and 20 photon cases. For the two-dimensional state space case the state required to be produced for the gate teleportation can be written as $\ket{\psi_{\text{prep}}}_{2}=\sum_{k, l=0}^{p, q}c_{k, l}\ket{k}\ket{l}$.  In Fig. \ref{fig:Pati2DFun} we can see how well the approximation in Eq.~(\ref{eqn:functionpoly}) holds for the $(p,q)=(3,3)$, $(5,5)$ and $(10,10)$ photon cases.

The preparation states $\ket{\psi_{\text{prep}}}_{1}$ and $\ket{\psi_{\text{prep}}}_{2}$ can be constructed with an optimisation and machine learning algorithm specifically designed for photonic quantum information tasks, i.e. the Strawberry Fields quantum software package \cite{SFpaper}. In Fig. \ref{fig:1D2DSF} we show how well the optimisation routine can approximate the desired preparation states. The gates used to construct these states are Gaussian gates and the Kerr gate. Note that the gate sequences for the construction of both of these states contain operators in both $\hat{x}_{i}$ and the conjugate variable $\hat{p}_{i}$. This means the gate sequence generated with the optimisation routine \cite{SFpaper} cannot be directly used for gate teleportation. 

Provided that the state $\ket{\psi_{\text{prep}}}$ has been appropriately prepared, the selective inversion operator can be implemented via teleportation in a similar way to the method described for the cubic phase gate in Sec. \ref{sec:cubic}. To complete the Grover search algorithm, one needs to construct a compound search operator $\mathcal{C}$~\cite{Pati2000}, defined as
\begin{eqnarray}
\mathcal{C} = - \hat{I}_{x_i} {F}^{\dag} \hat{I}_{x_f} {F},
\end{eqnarray}
where  $\hat{I}_{x_i}$ is the projection operator to the initial state and $\hat{I}_{x_f}$ is the projection operator to the final (target) state. The target state can be selected with high probability with approximately $\sqrt{N}$ applications of $\mathcal{C}$. As an example, we consider
the implementation of the Grover search algorithm with two-dimensional temporal cluster states for the case when $N=4$, as shown in Fig. \ref{fig:GroverTemporal}. The details of the implementation process is given in Table \ref{tab:CV-Grover}. In this particular case we would require two applications of $\mathcal{C}$, and four applications of the selective inversion operator $\hat{I}_{x}$~\cite{Pati2000}.

\section{Conclusion}\label{sec:conclusion}

We discussed in detail the implementations of three important quantum algorithms using the two-dimensional temporal cluster states: Gaussian Boson Sampling,  
CV-IQP and the CV Grover's search algorithm. We reviewed and summarised the simplified graphical representation and generation of one-dimensional and two-dimensional CV cluster states, and the
implementation of basic Gaussian and non-Gaussian gates (phase shifter, squeezer, beam splitter, cubic phase gate, {\rm etc.}) by homodyne measurements
on the temporal cluster. 

For Gaussian Boson Sampling, only Gaussian unitaries are required. Although the implementation of the original Boson Sampling using temporal cluster states has been discussed in Ref. \cite{Alexander2017MBLO}, we emphasise that implementing Gaussian Boson Sampling using temporal cluster states shows advantages in an experimental realisation. The is because the states and unitaries for Gaussian Boson Sampling  are Gaussian, meaning the errors due to finite squeezing in the cluster states can be corrected \cite{Su2018EC}. This could lead to the implementation of Gaussian Boson Sampling with a large number of modes, and thus may potentially help to achieve quantum supremacy
\cite{Boixo2018, Harrow2017, Preskill2018, Vankov2018, Chen2018}. 

For the CV-IQP circuit, we had to introduce non-Gaussian gates. The fundamental non-Gaussian gate we discussed was the cubic phase gate. In the CV-IQP circuit, all unitaries commute. We thus rearranged the gates such that all controlled-Z gates were to the right of the squeezers, all displacements were directly before the homodyne detectors and the non-Gaussian gates in between. The advantage of this rearrangement is similar to that we obtained for the Gaussian Boson Sampling: before the non-Gaussian gates the states and gates are Gaussian and  thus the errors due to finite squeezing can be corrected. This would increase the number of modes we can prepare given the overall noise tolerance. 

For the CV Grover's search algorithm we also require non-Gaussian gates. In this case we need to implement a selective inversion operator that can be reduced to a state teleportation problem, which we show is ultimately equivalent to a sequence of higher-order quadrature phase gates. We consider implementing the inversion operator with both a single and two continuous variable qumodes. In both cases we explicitly consider the state that would be required for gate teleportation. This state is simulated with the Strawberry Fields quantum software package \cite{SFpaper}. \\


\noindent {\bf Acknowledgements:} We would like to thank Joshua Izaac for help with data visualisation and Juan Miguel Arrazola $\&$ Thomas Bromley for valuable, patient explanations regarding the Strawberry Fields software package. 


\end{document}